\documentclass[12pt]{article}
\usepackage{amstext}
\usepackage{amssymb}

\newcommand\RR{{\mathbb R}}

\def\d={\,:=\,}

\newcommand{\semdir}{\rtimes}

\font\frakten=eufm10
\newfam\frakfam
\textfont\frakfam=\frakten

\newtheorem{thm}{Theorem}[section]
\newtheorem{lemma}[thm]{Lemma}
\newtheorem{cor}[thm]{Corollary}

\newtheorem{Defn}[thm]{Definition}
\newtheorem{Ex}[thm]{Example}
\newtheorem{Rem}[thm]{Remark}
\newtheorem{Exs}[thm]{Examples}
\newtheorem{Rems}[thm]{Remarks}
\newtheorem{Defrem}[thm]{Definition and Remark}
\newtheorem{Remnt}[thm]{}

\newenvironment{rem}
 {\begin{Rem} \begin{rm}} {\end{rm} \hfill $\diamond$ \end{Rem}}
\newenvironment{rems}
 {\begin{Rems} \begin{rm}} {\end{rm} \hfill $\diamond$ \end{Rems}}

\newenvironment{prf} {{\bf Proof.}}{\hfill $\diamond$}

\def\be{\begin{equation}}
\def\ee{\end{equation}}
\def\bea{\begin{eqnarray}}
\def\eea{\end{eqnarray}}

\newcommand{\bzeta}{\mbox{\boldmath $\zeta$}}
\newcommand{\bgam}{\mbox{\boldmath $\gamma$}}

\def\text#1{{\rm #1}}
\newcommand{\bfrakg}{\mbox{\boldmath $\mathfrak g$}}
\newcommand{\h}{\mathfrak H}

\def\Ad#1{\mathop{{\rm Ad}_{{#1}}}}
\def\coAd#1{\mathop{{\rm Ad}_{{#1}}\dual}}
\def\dual{^{\sharp}}
\def\sinch{{\,\hbox{sinch}\,}}
\def\cosinch{{\,\hbox{cosinch}\,}}

\def\lb{\hbox{$\lambda$\hskip-5pt$\raise2pt\hbox{\char'040}$\hskip1.5pt}}

\newcommand{\p}{\mbox{${\mathcal P}^{\uparrow}_{+}(1,1)$}}
\newcommand{\affp}{\mbox{${\mathcal P}_{\rm Aff}(1,1)$}}
\newcommand{\gwh}{\mbox{${\mathcal G}_{\rm WH}$}}

\begin{document}

\begin{center}
{\LARGE \bf Plancherel Inversion as Unified \\ [3mm]
    Approach to Wavelet Transforms \\ [3mm]
             and Wigner Functions}\\[10mm]

\sc S. Twareque Ali\footnote{e-mail: 
stali@mathstat.concordia.ca}, 
Hartmut F\"uhr\footnote{e-mail: fuehr@gsf.de }
 {\rm and} Anna E. Krasowska\footnote{e-mail: 
 ankras@alcor.concordia.ca}\\[3mm]

\small{\em Department of Mathematics and Statistics, 
Concordia University,\\ Montr\'eal, Quebec, CANADA H4B 1R6\\
Institut f\"ur Biomathematik und Biometrie, GSF, \\ 85764 Neuherberg,
Deutschland }

\end{center}


\begin{abstract}
  We demonstrate that  the Plancherel transform for 
Type-I groups provides one with a natural, 
unified perspective for the generalized continuous 
wavelet transform, on the one hand, and  for a class of 
Wigner functions, on the other. The wavelet transform 
of a signal is an ${\rm L}^2$-function on an appropriately chosen 
group while the Wigner function is defined on a coadjoint 
orbit of the group and serves as an alternative characterization
of the signal, which is often used in practical applications. 
The Plancherel transform maps ${\rm L}^2$-functions on a group unitarily
to fields of Hilbert-Schmidt operators, indexed by 
unitary irreducible representations of the group. The wavelet transform 
can essentially be looked upon as a restricted inverse Plancherel 
transform, while Wigner functions are modified Fourier transforms 
of inverse Plancherel transforms, usually restricted to a subset 
of the unitary dual of the group. Some known results on both 
Wigner functions and wavelet transforms, appearing in the literature
from very different perspectives, are naturally unified within 
our approach. Explicit computations on a number of groups 
illustrate the theory.
\end{abstract}

\section{Introduction}\label{sec:Intro}

    The continuous wavelet transform is used extensively 
in image processing and signal analysis and its group theoretical
origin is well known \cite{AAG-book,Da10}. The Wigner function 
has also been employed in the analysis of signals as well as in 
numerous quantum optical and quantum statistical computations
\cite{Ba,BeBe,HiOCWi,Wig,Wo}. It has been argued before that 
both the wavelet transform and the Wigner function owe their 
origin to the square integrability of certain group representations.
This point was discussed extensively in \cite{AlAtChWo}, where 
the square integrability of a single unitary irreducible 
representation of a group was exploited to build both a 
generalized wavelet transform and a class of Wigner functions.
It is the purpose of this paper to show, quite generally, how 
both these concepts can be unified using the Plancherel 
transform for Type-I groups. The Plancherel transform sets up 
a unitary isomprophism between the Hilbert space of square
integrable (with respect to the Haar measure) functions on the 
group and the direct integral Hilbert space (with respect to 
the Plancherel measure) built out of the 
spaces of Hilbert-Schmidt operators on the Hilbert spaces of 
unitary irreducible representations of the group. It is the inverse 
of this unitary map which, when appropriately restricted, leads 
to a generalized wavelet transform. On the other hand, taking the 
inverse Plancherel transform and following it up with a 
Fourier type of transform leads to functions on the dual of the 
Lie algebra of the group. The final function, when restricted to 
appropriate coadjoint orbits, then yields a wide class of 
generalized Wigner functions, which share many of the interesting 
properties of the original Wigner function \cite{Wig}, but now 
is definable for a vast array of groups and representations. 
Generalized wavelet transforms can also be seen as coherent 
state transforms of vectors in the Hilbert spaces of group 
representations \cite{AAG-book}. However, in most cases one 
defines the coherent state transform on Hilbert spaces carrying 
a single unitary irreducible representation of the group. This 
requires that the representation in question be square integrable, 
or in other words, that it belong to the discrete series of 
representations of the group. If, on the other hand, the group 
in question does not admit square integrable representations, the 
above construction of coherent states and related transforms 
clearly fails. In such cases, in specific examples, it has 
been demonstrated \cite{KlSt} how the  use of direct integral 
representations, over some convenient subset of the unitary dual 
of the group, leads once more to the existence of a coherent state 
transform. We show here that this situation is generic and is 
again a simple consequence of the Plancherel transform. 

   The rest of this paper is organized as follows: In 
Section \ref{sec-planchmeas} we briefly recall the Plancherel 
transform and its inverse for Type-I groups. In 
Section \ref{sec-inv_pl} we derive explicit expressions 
for the inverse Plancherel transform and demonstrate how 
it can be used 
to define coherent states and a generalized wavelet transform. 
We carry out the construction explicitly in 
Section \ref{sec-expoincare} for the case of the Poincar\'e 
group in a two-dimensional space-time. Section \ref{sec-wigfcn} is 
devoted to a definition and construction of the generalized Wigner 
function. This function is defined on the coadjoint orbits, 
foliating the dual of the Lie algebra of the group, and we 
introduce a modified Fourier transform on the range of the 
Plancherel transform to arrive at it. We also discuss general 
properties of the Wigner function, which follow immediately 
from the definition. As examples, we compute in 
Section \ref{sec-examples}
Wigner functions for the cases of three
commonly used groups: the two-dimensional Poincar\'e group, the
affine Poincar\'e group (the Poincar\'e group including 
dilations) and the Weyl-Heisenberg group, which leads us 
back to the original 
quasi-probability distribution function introduced by Wigner. 
Finally, in the Appendix we collect together a few results, of a 
computational nature, used in working out the examples.

\section{Plancherel Measure}\label{sec-planchmeas}

Let us first fix some notation: $G$ denotes a second countable, 
locally
compact group. All representations will be understood to be unitary
and strongly continuous. By $\widehat{G}$ we denote the set of 
equivalence classes of irreducible representations of $G$, 
equipped with the Mackey Borel structure(see, e.g., \cite{Fo}).
It will often be necessary to distinguish between an equivalence 
class 
$\sigma \in \widehat{G}$ of representations and a specific 
realization of a representation $U_{\sigma}$, in this equivalence
class and acting on a particular Hilbert space 
${\mathfrak H}_{\sigma}$. 
In the direct integrals below, a measurable realization of the 
representations $U_{\sigma}$ used is provided by the theory \cite[Theorem 
10.2]{Ma3}.
$\mu_G$ denotes the left Haar measure, and ${\rm L}^p(G)$ is the
corresponding ${\rm L}^p$-space. $C_c(G)$ denotes the space of
compactly supported continuous functions on $G$. The modular 
function of $G$ is
denoted by $\Delta_G$, the convention being, 
\begin{equation} 
d\mu_{G}(x) = \Delta_{G}(x)\; d\mu_{r}(x)~~, \label{eq:modfcn}
\end{equation}
where $\mu_{r}$ is the right invariant Haar measure. 
For a function $f$ on $G$, we write
$$\widetilde{f} (x) := \overline{f(x^{-1})}. $$

   For a given Hilbert space ${\mathfrak H}$, 
${\mathcal B}_2({\mathfrak H})$
denotes the space of Hilbert-Schmidt operators. It is a Hilbert
space, endowed with the scalar product $\langle A | B \rangle_2 
 = {\rm tr}(A^{*} B)$; the corresponding norm shall be 
denoted by $\| \cdot \|_2$.
Furthermore, ${\mathcal B}_1({\mathfrak H})$ denotes the subspace of
trace class operators, endowed with the norm 
$\| A \|_1 := {\rm tr} (|A|)$,
where  $|A| := (AA^*)^{1/2}$. Elements of special interest in both
spaces are the rank-one operators, denoted by 
$| \eta \rangle \langle
\phi |$, (for $\eta, \phi \in {\mathfrak H}$), which are defined by
$| \eta \rangle \langle \phi | (\psi) = 
\langle \phi | \psi \rangle \eta$, for any $\psi \in \mathfrak H$.
We have $\| | \eta \rangle \langle \phi | \|_1 
= \| | \eta \rangle \langle
\phi | \|_2 = \| \eta \| \| \phi \|$.

The usual operator norm is
denoted by $\| \cdot \|_{\infty}$. If a densely 
defined operator $A$ has a 
bounded extension, we denote the extension by $[A]$. A simple
and often used fact is that for linear operators $A,B,T$ with
$A,B$ bounded, such that $[AT]$ and $[TB]$ exist, 
$[ATB] = [AT] B = A [TB]$. 

The central object of interest in this paper is the {\em left
regular representation} $\lambda_G$ of $G$, acting on ${\rm L}^2(G)$
via $(\lambda_G(x) f)(y) := f(x^{-1}y)$. 
Another representation acting
on ${\rm L}^2(G)$ is the {\bf right regular representation} 
$\rho_G$,
defined by $(\rho_G(x) f)(y) := \Delta_G(x)^{1/2} f(yx)$. The left 
and the right regular representations commute and are unitarily 
equivalent.  Finally, the
{\bf two-sided representation} is denoted by 
$\lambda_G \times
\rho_G$. This is a representation of the product group $G \times G$,
defined by $(\lambda_G \times \rho_G) (x,y) 
:= \lambda_G(x) \rho_G(y)$.
The Plancherel theory can be seen as the theory of a direct integral
decomposition of the two-sided representation into irreducibles, 
where the
intertwining operator is given by the operator-valued Fourier 
transform. In this paper, we shall only be concerned with 
groups $G$ such that $\lambda_G$ is a Type-I factor,  i.e.,
the von Neumann algebra
generated by the left (right) regular representation \cite{DiW,Di}  of $G$
is a Type-I factor. (Such von Neumann algebras are algebraically
isomorphic, as $C^*$-algebras, to the full algebra of 
bounded operators on some Hilbert space.)

Recall that the operator-valued
Fourier transform on $G$ maps each $f \in {\rm L}^1(G)$ to the
family $\{U_{\sigma}(f)\}_{\sigma \in \widehat{G}}$ of operators,
where each $U_{\sigma}(f)$ is defined by the weak operator integral
\begin{equation} 
U_{\sigma}(f) := \int_G f(x) U_{\sigma}(x) d\mu_{G}(x) ~~.
\label{eq:opFourtrans}
\end{equation}
This defines a field of bounded operators, in fact, we have
\begin{equation} \label{ineq:NormFourier}
 \| U_{\sigma}(f) \|_{\infty} \le \| f \|_1
\end{equation}
Another feature of
the operator-valued Fourier transform, reminiscent of the well-known
Fourier transform over the reals, is that convolution becomes
operator multiplication on the Fourier side, more precisely,
$U_{\sigma}(f \ast g) = U_{\sigma} (f) \circ U_{\sigma} (g)$.
In order to invert this transform, we  have to find 
a Hilbert space ${\mathfrak H}$,
such that $f \mapsto \{U_{\sigma}(f)\}_{\sigma \in \widehat{G}}$
extends from ${\rm L}^1(G) \cap {\rm L}^2(G)$
to a unitary equivalence ${\rm L}^2(G) \to {\mathfrak H}$.
To see the relationship of this definition to the usual 
Fourier transform
(over the reals, say), let us suppose for a moment that $G$ 
is abelian.
Then each $U_{\sigma}(f)$ is a scalar, since each
$U_{\sigma}$ is a character, and the above mapping yields the 
usual Fourier
transform $\widehat{f}$ (except that in the generally used definition 
one integrates over $U_{\sigma}^*$ rather than $U_{\sigma}$). 
Also, in this case,
$\widehat{G}$ is a locally compact 
abelian group, and the abelian 
Plancherel theorem states that we may take the Haar measure 
on $\widehat{G}$ as the Plancherel measure, 
i.e. ${\mathfrak H} = {\rm L}^2(\widehat{G})$, in the stated unitary 
equivalence.

Returning to the general case, let us try to motivate the 
construction
of the Hilbert space ${\mathfrak H}$. The Fourier transform 
$\{U_{\sigma}(f)\}_{\sigma \in \widehat{G}}$
forms a field of bounded operators on $\widehat{G}$. Furthermore, 
this field is measurable, as follows from the definition 
of the Mackey
Borel structure on $\widehat{G}$. It is thus reasonable to 
expect ${\mathfrak H}$
to be the direct integral over a measure space 
$(\widehat{G},\nu_G)$, where each
fibre is some Hilbert space of operators, and the measure 
$\nu_G$ is
to be determined. The natural choice for the fibres is given by the
Hilbert-Schmidt operators on the representation spaces 
$\mathfrak H_{\sigma}$. 
At this point, the
Plancherel theory splits into the unimodular and the nonunimodular
cases: In the unimodular case, $U_{\sigma}(f)$ is automatically
Hilbert-Schmidt, for every $f \in {\rm L}^1(G) \cap {\rm L}^2(G)$
and almost every $\sigma \in \widehat{G}$. In the 
nonunimodular case
we have to employ a family $(C_{\sigma})_{\sigma \in \widehat{G}}$ 
of
densely defined unbounded operators $C_{\sigma}$ on 
${\mathfrak H}_{\sigma}$,
with densely defined inverses,
such that $U_{\sigma}(f) C_{\sigma}^{-1}$ is Hilbert-Schmidt
(more precisely: for almost all $\sigma$ (with
respect to the measure $\nu_{G}$), the closure  
$[U_{\sigma}(f) C_{\sigma}^{-1}]$ is Hilbert-Schmidt). 
These operators can indeed be constructed in such way 
that the operator Fourier
transform extends to a unitary map.

Let us now give the exact statement of the Plancherel theorem
in the form we are going to use \cite{DuMo}.

\begin{thm}\label{Pl-Thm}
 Let $G$ be a second countable locally compact group having a type-I
regular representation. Then there exists a measure 
$\nu_G$ on $\widehat{G}$, called the {\bf Plancherel measure}, 
and a measurable field $(C_{\sigma})_{\sigma \in \widehat{G}}$ 
of self adjoint
positive operators with densely defined inverses, 
such that the following hold:
\begin{enumerate}
\item[(i)] For $f \in {\rm L}^1(G) \cap {\rm L}^2(G)$ and
 $\nu_G$-almost all $\sigma \in \widehat{G}$, the closure 
of the operator
$U_{\sigma}(f)C_{\sigma}^{-1}$ is a Hilbert-Schmidt operator on
 ${\mathfrak H}_{\sigma}$. 
\item[(ii)] The map ${\rm L}^1(G) \cap {\rm L}^2(G) \ni f \mapsto
 \{[U_{\sigma}(f)C_{\sigma}^{-1}]\}_{\sigma \in \widehat{G}}$ extends
 to a unitary equivalence 
\begin{equation}
{\mathcal P}: {\rm L}^2(G) \to 
\int^{\oplus}_{\widehat{G}} {\mathcal B}_2({\mathfrak H}_{\sigma}) 
  d\nu_G(\sigma) ~~. 
\label{eq:planchtrans}
\end{equation}
 This unitary operator is called the {\bf Plancherel transform} 
of $G$. It has the intertwining property
\[ {\cal P} (\lambda_G(x) \rho_G(y) f) (\sigma) = 
 U_{\sigma}(x) \left(  {\cal P} (f) (\sigma) \right) U_{\sigma}(y)^* ~~.\]
\item[(iii)] There exists a subspace ${\mathcal D}(G) \subset
 {\rm L}^1(G) \cap {\rm L}^2(G)$, dense in ${\rm L}^2(G)$, 
such that for all
$f \in {\mathcal D}(G)$ and $\nu_G$-almost all $\sigma
 \in \widehat{G}$, the operator 
\[ [U_{\sigma}(f) C_{\sigma}^{-2}]
  = [[U_{\sigma}(f) C_{\sigma}^{-1}]C_{\sigma}^{-1}]\] 
is
densely defined and has a trace class extension, and
we have the Fourier inversion formula,
\begin{equation}
f(x) = \int_{\widehat{G}} {\rm tr}
 \left([U_{\sigma}(x)^* U_{\sigma}(f) C_{\sigma}^{-2}]
 \right) d\nu_G(\sigma) ~~. 
\label{eq:planchinverse}
\end{equation}
\item[(iv)] The Plancherel measure is essentially unique: 
The {\bf covariance relation} 
\begin{equation} \label{eq:cov_relation}
 U_{\sigma}(x) C_{\sigma} U_{\sigma}(x)^* = \Delta_G(x)^{1/2} C_{\sigma}
\end{equation}
fixes each $C_{\sigma}$ uniquely up to multiplication by a scalar, 
and once these are fixed, so is $\nu_G$. 
 Conversely, one can fix $\nu_G$ (which is a priori only unique up
 to equivalence) and thereby determine the $C_{\sigma}$ uniquely.
\item[(v)] $G$ is unimodular if and only if for 
$\nu_G$-almost all $\sigma$,  $C_{\sigma}$ is 
a multiple of the identity $I_{\sigma}$ 
on ${\mathfrak H}_{\sigma}$.
 In this case we require that $C_{\sigma} = I_{\sigma}$, 
which then determines $\nu_G$ completely.  If $G$ 
is nonunimodular, $C_\sigma$ is an unbounded operator 
for ($\nu_G$-almost all) $\sigma \in \widehat{G}$.
\end{enumerate}
\end{thm} 

\begin{rem}
The inversion formula (\ref{eq:planchinverse})
was shown in \cite{DuMo} to hold for
the space of Bruhat functions introduced in \cite{Br}. 
It can be written as
\[ {\mathcal D}(G) = \bigcup \{ C_c^{\infty}(G/K) : K \subset G
 \mbox{ compact such that $G/K$ is a Lie group }~\} ~~,\]
where $C_c^{\infty}(G/K)$ is the space of arbitrarily smooth
functions on $G/K$ with compact support, 
canonically embedded into $C_c(G)$. 
\end{rem}

In the following, we suppose that $G$ is a second 
countable group with
type-I regular representation. We use $\widehat{\;}$ to denote the
Plancherel transform. So, for 
$f \in {\rm L}^1(G) \cap {\rm L}^2(G)$ we have 
$({\mathcal P}f)(\sigma ) = \widehat{f}(\sigma) 
:= [U_{\sigma}(f) C_{\sigma}^{-1}]$.
The direct integral space of Hilbert-Schmidt spaces in 
(\ref{eq:planchtrans}) will
be denoted by ${\mathcal B}_2^{\oplus}$. The scalar product 
of two elements, $A^i \in {\mathcal B}_2^{\oplus}, \;\; i = 1,2$, 
consisting of the measurable fields $\{ A^i (\sigma ) \in 
{\mathcal B}_2 (\h_\sigma ) \}_{\sigma \in \widehat{G}}$, is 
given by
$$ \langle A^1 \vert A^2 \rangle_{{\mathcal B}_2^{\oplus}} 
   = \int_{\widehat{G}} \mbox{\rm tr}\; [A^1 (\sigma )^*
    A^2 (\sigma ) ] \; d\nu_G (\sigma ) . $$ 

\section{The wavelet transform as inverse Planche\-rel transform}
\label{sec-inv_pl}

Let us quickly recall the group-theoretical formalism for the
construction of wavelet transforms: Suppose we are given a
unitary (not necessarily irreducible) representation $U$ of $G$, 
on a Hilbert space $\mathfrak H$ and a vector
$\eta \in {\mathfrak H}$. We can then define the 
(generalized) {\bf wavelet
transform} of $\phi \in {\mathfrak H}$ as the function 
$V_{\eta} \phi$ on $G$, defined by
\begin{equation}
 (V_{\eta} \phi )(x) := \langle  U(x) \eta | \phi \rangle ~~.
\label{eq:wavelettransf}
\end{equation}
Generally, this construction gives an injective operator
$V_{\eta}: \phi \mapsto V_{\eta} \phi$,
whenever $\eta$ is a cyclic vector (which means that the orbit
$U(G) \eta$ is total in ${\mathfrak H}$). However, in order to have
an efficient way of inverting $V_{\eta}$, we require 
more, {\em viz\/},
that $V_{\eta} : {\mathfrak H} \to {\rm L}^2 (G)$ be an isometry
(possibly up to a scalar factor $c_{\eta}$). Note that generally,  even 
the well-definedness,
that is $V_{\eta}({\mathfrak H}) \subset {\rm L}^2(G)$, is not
guaranteed.
However, if $V_\eta$ is an isometry (in which case we say that
$\eta$ is an {\bf admissible 
vector}), we can rewrite the isometry property in the form of 
an inversion formula,
\begin{equation}
 \phi = \frac{1}{c_{\eta}}
 \int_G (V_{\eta} \phi)(x) \;U(x) \eta \;d\mu_G (x) ~~,
\label{eq:invisom}
\end{equation}
where the integral is understood in the weak operator sense or,
equivalently, as a resolution of the identity:
\begin{equation}
{\rm Id}_{{\mathfrak H}} =  \frac{1}{c_{\eta}}
   \int_G  U(x) |\eta \rangle \langle \eta | U(x)^*\;
 d\mu_G (x) ~~.
\label{eq:resolid1}
\end{equation}

The relationship to the left regular representation is quite 
obvious: Besides being an isometry,
the wavelet transform $V_{\eta}$ is easily seen to intertwine
the representation $U$ of $G$ with its left regular 
representation $\lambda_G$. Hence wavelet transforms 
fall quite 
naturally in the domain of the Plancherel theory. In fact, as will 
become
clear below, if $U$ is given as a direct integral of irreducible
representations, a wavelet transform is just the inverse 
Plancherel transform, applied to certain operator fields.

In order to motivate the last statement, let us take a closer
look at the case where $U = U_{\sigma}$ is 
an irreducible, {\bf square-integrable
representation} of $G$. This means that, there exist explicit 
admissibility
conditions involving $C_{\sigma}$, 
which are usually cited in the following
form \cite[Theorem 3]{DuMo}:
\begin{thm}
\label{AdmCond}
 Let $U_{\sigma}$ be an irreducible subrepresentation of 
$\lambda_G$.
\begin{enumerate}
 \item[(i)] For $\eta, \phi \in {\mathfrak H}_{\sigma}$, 
the wavelet transform
 $V_{\eta} \phi$ is square integrable (i.e., is an element of 
${\rm L}^{2}(G)$) iff $\eta \in {\rm dom} (C_{\sigma})$.
 \item[(ii)] For $\eta_1,\eta_2 \in {\rm dom}(C_{\sigma})$ and 
$\phi_1,
 \phi_2 \in {\mathfrak H}_{\sigma}$, we have the orthogonality 
relation,
 \begin{equation}
 \langle V_{\eta_1} \phi_1 | V_{\eta_2} 
  \phi_2 \rangle_{{\rm L}^{2}(G)} = 
 \langle C_{\sigma} \eta_2 | C_{\sigma} \eta_1 \rangle_{\mathfrak H}
  \;\langle \phi_1 | \phi_2 \rangle_{\mathfrak H} ~~.
 \label{eq:orthogonality1}
\end{equation}
 \end{enumerate}
\end{thm}

To see the relationship with the  Plancherel transform, let us 
consider the
rank-one operators $A_i = | \phi_i \rangle \langle
C_{\sigma} \eta_i |$
($i=1,2$). Then $V_{\eta_i} \phi_i (x) = 
\langle U_{\sigma}(x) \eta_i | \phi_i \rangle = 
{\rm tr} (| \phi_i \rangle
 \langle \eta_i | U_{\sigma}(x)^*) = {\rm tr} ( | \phi_i \rangle
 \langle C_{\sigma} \eta_i |C_{\sigma}^{-1} U_{\sigma}(x)^*) 
= {\rm tr}\;[ A_i C_{\sigma}^{-1} U_{\sigma}(x)^* ]$,
which is essentially the Plancherel inversion formula 
(\ref{eq:planchinverse})
(up to an ordering of operators),
with the operator fields supported only at the point $\sigma$.
Here we have taken account of the fact that 
$U_{\sigma}$ is a subrepresentation of
$\lambda_G$ iff $\nu_G(\{ \sigma \}) > 0$, and hence, without 
loss of generality, we may take $\nu_G (\{ \sigma \})= 1$.
Assuming that the inversion formula holds for both 
$V_{\eta_i} \phi_i, \;\; i=1, 2$,
we obtain $({\mathcal P}(V_{\eta_i} \phi_i)) (\pi) = 
\widehat{(V_{\eta_i} \phi_i )}(\pi ) = A_i$, for $\pi = 
\sigma$ and $0$ elsewhere. Thus,
the orthogonality relations, and in particular the isometry 
property of the generalized wavelet transform $V_{\eta}$, 
are immediate consequences of the unitarity of 
the Plancherel transform. 

While this way of showing the isometry property 
of $V_{\eta}$, using the Planche\-rel transform, is 
much too complicated 
in the irreducible 
case (which is easily dealt with using Schur's lemma), 
it has the advantage of
being readily generalizable to direct integral representations, 
once we have
extended the inversion formula 
(\ref{eq:planchinverse}) to a wider class of functions.

Let us first establish a few preliminary facts. The first lemma deals
with the operators $C_{\sigma}$ and their relation to convolution.
\begin{lemma}
\label{convlemma}
 Let $f \in C_c(G)$. 
 \begin{enumerate}
 \item[(i)] For $\nu_G$-almost every $\sigma$, we have 
  $\widehat{f}(\sigma)^* = C_{\sigma}^{-1} 
  U_{\sigma}(\Delta_G^{-1} \widetilde{f})$.
  In particular the right hand side is everywhere 
defined and bounded.
 \item[(ii)] For $\nu_G$-almost every $\sigma$, we have  
 \[ [U_{\sigma}(f) C_{\sigma}^{-1}]  = C_{\sigma}^{-1} 
  U_{\sigma}(\Delta_G^{-1/2} f) 
 ~~,\]
  in particular the right hand side is everywhere defined 
and bounded.
 \item[(iii)] For all $g \in {\rm L}^2(G)$, we have 
 \begin{eqnarray*}
  (\widehat{g \ast f}) (\sigma) & = & \widehat{g}(\sigma) 
   U_{\sigma}(\Delta_G^{-1/2} f)
  ~~, \\
  (\widehat{f \ast g}) (\sigma) & = & U_{\sigma}(f) 
\widehat{g}(\sigma) ~~.
 \end{eqnarray*}
 \end{enumerate}
\end{lemma}

\begin{prf}
 For part $(i)$ we invoke \cite[Theorem 13.2]{Ru}, to find that, 
since
 $C_{\sigma}^{-1}$ is self-adjoint and $U_{\sigma}(f)$ is bounded,
 $(U_{\sigma}(f) C_{\sigma}^{-1})^* = C_{\sigma}^{-1} U_{\sigma}(f)^*$.
 Moreover, since $\widehat{f}(\sigma )$ is bounded, 
the right hand side of the last
 equation is everywhere defined. Calculating 
$U_{\sigma}(f)^*$ is routine.
 
 For $(ii)$ we first note that by $(i)$, applied to $\Delta_G^{-1/2}
 \widetilde{f} \in C_c(G)$,  the right hand side is bounded and
 everywhere defined. Moreover, the left-hand side is bounded since
 $f \in {\rm L}^1(G) \cap {\rm L}^2(G)$. It thus remains to show
 that the equality holds on the dense subspace 
${\rm dom} (C_{\sigma}^{-1})$:
 For $\phi, \eta
 \in {\rm dom}(C_{\sigma}^{-1})$ the definition of the weak operator
 integral yields
 \begin{eqnarray*} \langle \phi | U_{\sigma}(f) C_{\sigma}^{-1} \eta
 \rangle & = & \int_G \langle \phi | U_{\sigma}(x) C_{\sigma}^{-1} \eta
 \rangle f(x) d\mu_G(x) \\
 & = & \int_G \langle \phi | \Delta_G(x)^{-1/2} C_{\sigma}^{-1} U_{\sigma}(x)
 \eta \langle f(x) d\mu_G(x) \\
 & = &  \int_G \langle C_{\sigma}^{-1} \phi | U_{\sigma}(x)
 \eta \langle  \Delta_G(x)^{-1/2} f(x) d\mu_G(x) \\
 & = & \langle  C_{\sigma}^{-1} \phi | U_{\sigma}(\Delta_G^{-1/2} f) \eta
 \rangle \\
 & = &  \langle \phi |  C_{\sigma}^{-1} U_{\sigma}(\Delta_G^{-1/2} f) \eta
 \rangle ~~,
 \end{eqnarray*}
 where the second equality uses the covariance relation
 (\ref{eq:cov_relation}), and the self-adjointness of $C_{\sigma}^{-1}$
 was used on various occasions. This shows $(ii)$.

 Part $(iii)$ is then immediate from $(i)$ and $(ii)$, at least for
 $g \in {\rm L}^1(G) \cap {\rm L}^2(G)$. It extends by continuity to
 all of ${\rm L}^2(G)$: The left-hand sides are continuous
 operators, being convolution operators with $f \in C_c(G)$, and the right
 hand sides are continuous because of inequality (\ref{ineq:NormFourier}).
\end{prf}

The next lemma defines the space ${\mathcal B}_1^{\oplus}$, which arises
very naturally when dealing with inversion formulae. In fact, there
is a natural representation-theoretic interpretation of ${\mathcal
B}_1^{\oplus}$ as the space of Fourier transforms of the Fourier
algebra $A(G)$. This was noted for the unimodular case by Lipsman
\cite{Li}, but the arguments go through for the non-unimodular
case as well.

\begin{lemma}
 Let ${\mathcal B}_1^{\oplus}$ be the space of measurable fields
 $\{B(\sigma)\}_{\sigma \in \widehat{G}}$ of trace class operators, 
for which the norm
 \[ \left\|B
  \right\|_{{\mathcal B}_1^{\oplus}} :=
    \int_{\widehat{G}} \left\|B(\sigma ) \right\|_1 d\nu_G(\sigma) 
 \]
 is finite. Here we identify operator fields which agree 
 $\nu_G$-almost everywhere. Then $({\mathcal B}_1^{\oplus},\left\| 
  \cdot \right\|_{{\mathcal B}_1^{\oplus}})$
 is a Banach space and the set of measurable fields of rank 
one operators
 in ${\mathcal B}_1^{\oplus}$ spans a dense subspace.
\end{lemma}
The proof consists of standard arguments and is omitted here.

Now we can show that the inversion formula holds almost everywhere, 
whenever it makes sense (i.e., whenever all quantities 
involved can be expected
to converge). This is the nonabelian analogue of the well known
Fourier inversion formula for an ${\rm L}^2$-function whose
Plancherel transform is in ${\rm L}^1$.
The statement for the unimodular case was in fact given in \cite{Li}.

\begin{thm}
\label{Pl-Inv}
 Let $A \in {\mathcal B}_2^{\oplus}$ be 
 such that for almost all $\sigma \in \widehat{G}$,  
$A(\sigma) C_{\sigma}^{-1}$ 
 extends to a trace class operator and such that 
 $\{[A(\sigma) C_{\sigma}^{-1}]\}_{\sigma \in \widehat{G}} \in
 {\mathcal B}_1^{\oplus}$.
 Let $a \in {\rm L}^2(G)$ be the inverse Plancherel transform of $A$.
 Then we have (for almost every $x \in G$)
\be a(x) = \int_{\widehat{G}} 
  {\rm tr}(U_{\sigma}(x)^* [A(\sigma)C_{\sigma}^{-1}])
 \; d\nu_G(\sigma)~~.
\label{eq:invexp}
\ee
 If we assume that $C_{\sigma}^{-1} A(\sigma)$ is trace-class, 
and that $\{[C_{\sigma}^{-1} A(\sigma)]\}_{\sigma \in \widehat{G}} 
\in {\mathcal B}_1^{\oplus}$, then 
 $\sigma \mapsto {\rm tr}(|[C_{\sigma}^{-1} A(\sigma)]|)$
 is integrable and we obtain (almost everywhere)
\be
 a(x) = \int_{\widehat{G}} {\rm tr}([ C_{\sigma}^{-1}
 U_{\sigma}(x)^*A(\sigma)]) d\nu_G(\sigma)~~.
\label{eq:invexp2}
\ee

\end{thm}
 
\begin{prf}
 Let $(f_n)_{n \in N} \subset C_c(G)$ be a sequence with
 decreasing supports, satisfying the following requirements:
 $f_n \ge 0$, $\| \widetilde{f_n} \|_{\rm L^1} = 1$, 
 and ${\rm supp} (f_n)$ runs through a neighborhood base at unity.
 Then $(f_n)_{n \in N}$ is a bounded approximate identity
 with respect to right convolution, i.e., for all 
$g \in {\rm L}^2(G)$
 we have $g \ast f_n \to g$ in ${\rm L}^2(G)$, and the operator norms of
 $g \mapsto g \ast f_n$ are bounded by a constant. 
 In addition, $(\Delta_G^{-1/2} f_n)_{n \in N}$ has the same 
properties,
 since $\| \Delta_G^{1/2} \widetilde{f_n} \|_{{\rm L}^1} \to 1$. 
 
 By passing to a subsequence, if necessary, we may assume
 that $a \ast f_n \to a$ pointwise almost everywhere.
 We first evaluate the convolution using the unitarity of the 
Plancherel
 transform, obtaining for almost every $x \in G$:
 \begin{eqnarray}
  (a \ast f_n)(x) & = & \langle \lambda(x) \widetilde{f_n}
  | a \rangle  \nonumber \\
   & = & \int_{\widehat{G}} {\rm tr} \left(A(\sigma) 
    \left[U_{\sigma}(x) 
   U_{\sigma}(\widetilde{f_n}) C_{\sigma}^{-1}\right]^*\right)
   d\nu_G(\sigma) \label{ch_order} \\
   & = & \int_{\widehat{G}} {\rm tr} (A(\sigma)
   [ C_{\sigma}^{-1} U_{\sigma}(\widetilde{f_n})^*
   U_{\sigma}(x)^*] d\nu_G(\sigma) \nonumber \\
   & = & \int_{\widehat{G}} {\rm tr} ([A(\sigma) C_{\sigma}^{-1}
   U_{\sigma}(\Delta_G^{-1} f_n)  
   U_{\sigma}(x)^*]) d\nu_G(\sigma) ~~, \nonumber \\
   & = & \int_{\widehat{G}} {\rm tr} ([A(\sigma) C_{\sigma}^{-1}]
   U_{\sigma}(\Delta_G^{-1} f_n)  
   U_{\sigma}(x)^*) d\nu_G(\sigma) ~~. \label{eq:invexp_appr}
 \end{eqnarray}
 Here we have used the fact that $C_{\sigma}^{-1}
 U_{\sigma}(\Delta_G^{-1} f_n)  
 U_{\sigma}(x)^*$ is bounded, by Lemma \ref{convlemma}
 $(i)$,
 as well as the existence of $[A(\sigma) C_{\sigma}^{-1}]$, 
as assumed.

 From the definition of ${\cal B}_1^{\oplus}$, it is clear that 
 \[ (B(\sigma))_{\sigma \in \widehat{G}} \mapsto 
 \int_{\widehat{G}} {\rm tr} (B(\sigma)) d\nu_G(\sigma) 
 \]
 defines a bounded linear functional; this was our motivation for
 introducing the space. Comparing the right-hand
 side of (\ref{eq:invexp}) with (\ref{eq:invexp_appr}), we find that 
 it suffices to show that
 the sequence of operators 
 \begin{eqnarray*}
 T^{(n)}_1: {\mathcal B}_1^{\oplus}
 & \to & {\mathcal B}_1^{\oplus} \\
 ( B(\sigma))_{\sigma \in \widehat{G}} & \mapsto &
  \left( B(\sigma ) U_{\sigma}(\Delta_G^{-1} f_n) \right)_{\sigma \in
 \widehat{G}} \end{eqnarray*}
 converges strongly to the identity operator. For this purpose,
 let us write $T^{(n)}_2: {\mathcal B}_2^{\oplus} \to 
 {\mathcal B}_2^{\oplus}$ for the identically defined operators on
 ${\mathcal B}_2^{\oplus}$. Let us first note that 
 $\left\| U_{\sigma}(\Delta_G^{-1} f_n) \right\|_{\infty} \le \left\| 
 \Delta_G^{-1} f_n
 \right\|_{{\rm L}^1} \le K$, with $K$ independent
 of $n$, thus both sequences of operators are norm-bounded.

 Applying Lemma \ref{convlemma} (ii), we find that
 the Plancherel transform conjugates $(T^{(n)}_2)_{n \in N}$ with the 
 family of convolution operators $S^{(n)} : g \mapsto g \ast 
 (\Delta_G^{-1/2} f_n)$, which strongly converges to the identity operator.
 Moreover, it strongly converges with respect to the 
 ${\mathcal B}_1^{\oplus}$-norm 
 on the subspace generated by the fields of rank one operators:
 Let $B = \{ | \phi(\sigma) \rangle \langle 
  \eta(\sigma) | \}_{\sigma \in
 \widehat{G}}$ be such a field. We may assume $\left\|\phi(\sigma)\right\|
 = \left\| \eta(\sigma) \right\|$. Then $(T^{(n)}_1 B)(\sigma )=
 | \phi(\sigma) \rangle \langle 
 T^{(n)^*}_\sigma \eta(\sigma) | , \;\;
\sigma \in \widehat{G}$, with $T^{(n)}_{\sigma} =  
 U_{\sigma}(\Delta_G^{-1} f_n)$.
 Hence
 \begin{eqnarray*}
  \left\| T^{(n)}_1 B - B \right\|_{{\mathcal B}_1^{\oplus}} & = & 
 \int_{\widehat{G}} \left\| \phi(\sigma) \right\| 
\left\|  T^{(n)^*}_\sigma \eta(\sigma)
 - \eta(\sigma) \right\| d\nu_G(\sigma) \\
 & \le& \left( \int_{\widehat{G}} \left\| \phi(\sigma) \right\|^2
  d\nu_G(\sigma) \right)^{1/2}\\
 &\times&\left( \int_{\widehat{G}} \left\|  T^{(n)^*}_\sigma
  \eta(\sigma) - \eta(\sigma) \right\|^2 d\nu_G(\sigma) \right)^{1/2} ~~,
 \end{eqnarray*}
 by the Cauchy-Schwarz inequality. Here we have used that
 $\| | \phi(\sigma) \rangle \langle \eta(\sigma) | \|_1 = 
\| \phi(\sigma) \|
 \| \eta(\sigma) \| = \| \phi(\sigma)\|^2 = 
\| \eta(\sigma) \|^2$, hence
 all integrals converge. Picking any measurable family 
$\{\xi(\sigma)\}_{\sigma
 \in \widehat{G}}$ of unit vectors, we can define the operator field
 $ B'= \{ | \xi(\sigma) \rangle \langle 
\eta(\sigma) | \}_{\sigma \in \widehat{G}}
 \in {\mathcal B}_2^{\oplus}$, and find that
 \[ \int_{\widehat{G}} \left\|  T^{(n)^*}_\sigma
  \eta(\sigma) - \eta(\sigma) \right\|^2 d\nu_G(\sigma) =
  \left\| T^{(n)}_2 B' - B' \right\|_{{\mathcal B}_2^{\oplus}}^2 \]
 converges to zero. 

 Thus $(T_1^{(n)})_{n \in N}$ is a bounded sequence of operators 
 converging strongly on a dense subspace, which entails strong
 convergence on ${\mathcal B}_1^{\oplus}$. Hence, 
 \begin{eqnarray*}
 \lefteqn{{\rm lim}_{n \to \infty} \int_{\widehat{G}} {\rm tr}
             ([A(\sigma) C_{\sigma}^{-1}]
    U_{\sigma}(\Delta_G^{-1} f_n) U_{\sigma}(x)^*) d\nu_G(\sigma)
        = } \\ 
        & & {}\qquad \qquad \int_{\widehat{G}} {\rm tr} ([A(\sigma) 
  C_{\sigma}^{-1}] U_{\sigma}(x)^*)
             d\nu_G(\sigma) ~~, 
\end{eqnarray*}
 and the first equation is proved.

 The second formula is proved by modifying the argument 
for the first:
 We employ
  \[ {\rm tr} \left(A(\sigma) \left[U_{\sigma}(x) 
   U_{\sigma}(\widetilde{f_n})
  C_{\sigma}^{-1}\right]^*\right) = 
 {\rm tr} \left( \left[U_{\sigma}(x) 
U_{\sigma}(\widetilde{f_n}) C_{\sigma}^{-1}\right]^*A(\sigma)
\right) \]
in equation (\ref{ch_order}).

 After using
\[ [C_{\sigma}^{-1} U_{\sigma}(\Delta_G^{-1} f_n) 
U_{\sigma}(x)^* A(\sigma)] =
U_{\sigma}(\Delta_G^{-1/2} f_n) [C_{\sigma}^{-1} 
U_{\sigma}(x)^* A(\sigma)] ~~,\]
the fact that $(\Delta_G^{-1/2} f_n)_{n \in N}$ 
is a bounded approximate
identity with respect to left convolution now gives the desired 
convergence of the traces, and we are done.

\end{prf}

\begin{rem}
Before we apply the theorem to general direct integral 
representations,
let us first consider the relevance of the two inversion formulae,
(\ref{eq:invexp}) and (\ref{eq:invexp2}), for the irreducible case.
So let $\pi \in \widehat{G}$, $U_\pi$ a representation in
this class and 
 assume it to be an irreducible subrepresentation of $\lambda_G$. Let
$\phi,\eta \in {\mathcal H}_{\pi}$ with $\eta \in {\rm dom}(C_{\pi})$.
The rank one operator $| \phi \rangle \langle C_{\pi} \eta|$ fulfills
the requirement for the first inversion formula, hence
$V_{\eta} \phi$ is the inverse Plancherel transform of this operator.
But we can also consider the operator $| C_{\pi} \eta \rangle 
\langle \phi |$,
suitable for the second inversion formula, which gives
\begin{eqnarray*} {\rm tr}(C_{\pi}^{-1} | U_{\pi}(x)^* 
C_{\pi} \eta 
\rangle \langle \phi |) & = & \langle \phi | C_{\pi}^{-1} 
U_{\pi}(x)^*
C_{\pi} \eta \rangle \\
& = & \langle \phi | \Delta_G^{-1/2}(x) U_{\pi}(x)^* \eta \rangle =
 (\Delta_G^{-1/2}  \widetilde{V_{\eta} \phi}) (x) ~~.
\end{eqnarray*}
This reveals the general relationship between the two 
inversion formulae:
The operators $f \mapsto \Delta_G^{-1/2} \widetilde{f}$ and
$\{A(\sigma)\}_{\sigma \in \widehat{G}} \mapsto 
\{A(\sigma)^*\}_{\sigma \in \widehat{G}}$ are conjugate under 
the Plancherel
transform, hence an inversion formula for $f$ gives rise to an
inversion formula for $\Delta_G^{-1/2} \widetilde{f}$, 
and vice versa.
\end{rem}

Now let $U_\pi$ be a multiplicity-free subrepresentation
of $\lambda_G$. Since $G$ has a type-I regular representation, 
we may assume
\[ U_{\pi} = \int_{\Sigma}^{\oplus} U_\sigma d\nu_G (\sigma) ~~,\]
for some measurable subset $\Sigma \subset \widehat{G}$.
A simple method for the construction of admissible vectors 
is then given in the following corollary:

\begin{cor}
\label{cor:mult_free}
Let $\phi = \{\phi(\sigma)\}_{\sigma \in \Sigma},\;\;
\{\eta(\sigma)\}_{\sigma \in \Sigma} \in {\mathcal H}_{\pi}$ be given. 
Assume, moreover, that $\eta(\sigma) \in {\rm dom}(C_{\sigma})$,
and that the field $A(\sigma) := 
|\phi (\sigma )\rangle \langle C_{\sigma} \eta (\sigma) |$, extended
trivially outside $\Sigma$, is in ${\mathcal B}_2^{\oplus}$. 
Then $V_{\eta} \phi \in {\rm L}^2(G)$, with
$\widehat{(V_{\eta} \phi)} = A$, and hence 
\[ \left\| V_{\eta} \phi \right\|^2 =
\int_{\Sigma} \left\| \phi(\sigma) \right\|^2 \left\|
C_{\sigma} \eta(\sigma) \right\|^2 d\nu_G(\sigma) ~~.\]
Thus, $\eta$ is admissible iff 
$\{\eta (\sigma )\}_{\sigma \in \Sigma}$ can be chosen such that
$\| C_{\sigma} \eta(\sigma) \| = 1$,
for $\nu_G$-almost every $\sigma \in \Sigma$.
\end{cor}

\begin{prf}
Let $a \in {\rm L}^2(G)$ be the inverse Plancherel transform of $A$.
Then, observing that $A(\sigma) C_{\sigma}^{-1} = 
| \phi(\sigma) \rangle
\langle \eta(\sigma) |$, we see that
as a function of $\sigma$, 
  ${\rm tr} (|A(\sigma) C_{\sigma}^{-1}|) = \| \phi(\sigma) \|
\| \eta(\sigma) \|$ is
integrable, since $\phi$ and $\eta$ are square-inte\-grable 
vector fields.
Hence all requirements of Theorem \ref{Pl-Inv} are met, 
and we obtain
almost everywhere

 \begin{eqnarray*}
a (x) & = &  \int_{\widehat{G}} {\rm tr}(A(\sigma)C_{\sigma}^{-1}
U_{\sigma}(x)^*) d\nu_G(\sigma) \\
& = & \int_{\widehat{G}} \langle U_{\sigma}(x) \eta(\sigma) | 
\phi(\sigma) 
\rangle d\nu_G(\sigma) \\
& = &  (V_{\eta} \phi ) (x)
\end{eqnarray*}
The equality of norms is then immediate,
since the right hand side is the norm squared of 
$A$ in ${\mathcal B}_2^{\oplus}$;
and the admissibility condition is an immediate corollary.
\end{prf}

The construction of admissible vectors for representations
with multiplicities can be a subtle task. For instance, it
is known that $\lambda_G$ has admissible vectors, for $G$
non-unimodular with type-I regular representation \cite{Fu},
but a direct construction of such vectors, without the use
of Plancherel transform, could not be given. By contrast,
the admissibility condition of the corollary is fairly
easy to handle, once the direct integral decomposition
of the representation is obtained. One important class
of representations which fall under this category are the
quasi-regular representations of certain semidirect product
groups, see \cite{FuMa}.

\begin{rem}
At the moment, we do not know whether the requirement
$\eta_{\sigma}  \in
{\rm dom}(C_{\sigma})$ is necessary for admissibility, 
i.e., for the finiteness of $\left\| V_{\eta} \phi \right\|^2$,
though we expect it to be true.
\end{rem}

A criterion for the existence of admissible vectors is 
given in the following
theorem. The proof for the unimodular part is a
straightforward consequence of Corollary \ref{cor:mult_free}, for
the non-unimodular case see \cite{Fu}, where in fact
all representations having admissible vectors are classified.

\begin{thm}\label{th-admissvect}
The unitary representation $U_\pi$ has admissible 
vectors iff $G$ is nonunimodular or
$G$ is unimodular and $0 < \nu_G(\Sigma) < \infty$.
\end{thm}

\section{Example of the (1+1)-Poincar\'e group}\label{sec-expoincare}
In this section we want to calculate the Plancherel measure
of the Poincar\'e group in $(1+1)$-dimensional space-time. 
This is the group 
$\p =\RR^2 \semdir {\rm SO}_{0}(1,1)$ (connected part of 
${\rm SO}(1,1)$) and we shall explicitly construct admissible vectors for 
some of its representations. Note that ${\rm SO}_{0}(1,1)$ 
is the proper Lorentz 
group in a space-time of $(1+1)$-dimensions. In computing 
the Plancherel 
measure, we follow the
procedure given by Kleppner and Lipsman \cite{KlLi}, which
employs the Mackey machinery for this purpose. 
Recall that 
it follows from Mackey's theory of induced representations
\cite{Ma68,Ma2}, that (almost all of) the  
unitary irreducible representations of 
$\RR^2 \semdir {\rm SO}_{0}(1,1)$ are in 
one-to-one correspondence with the orbits of ${\rm SO}_{0}(1,1)$ in 
the dual space $\widehat{\mathbb R}^2$ (which we identify 
here with ${\RR}^2$ itself). Here we have used the fact 
that ${\rm SO}_0(1,1)$ operates freely on $\widehat{\RR^2} \setminus
 \{ 0 \}$, such that each dual orbit contributes precisely one
irreducible representation, and we have dropped the one-dimensional
representations arising from the dual orbit $\{ 0 \}$.

We parametrize the Lorentz group by 
\be 
\RR \ni \theta \mapsto \Lambda_\theta = \left( \begin{array}{cc} 
\cosh \theta & \sinh \theta \\ \sinh \theta & \cosh \theta
\end{array} \right) ~~. 
\label{lorentzmat}
\ee
In this parametrization, $d\theta$ is invariant, under both left 
and right actions, and we
choose this for the Haar measure on ${\rm SO}_{0}(1,1)$. We write a 
generic element of $G$ as $(x, h)$, with $x = \pmatrix{x_0 
\cr \mathbf x } \in 
{\RR}^2$ and $h$ a matrix of the form (\ref{lorentzmat}). As
Haar measure on $\p$ we may take $d\mu (x , h ) = dx d\theta$, 
$dx$ being the Lebesgue measure 
on ${\mathbb R}^2$, and note that this group is 
unimodular. 

  The first step for the calculation of the Plancherel
measure is the computation of the dual
orbits. They are conveniently represented by the set
\be \left\{ y v \;\vert\; v \in \{ \pmatrix{1 \cr 0}, \; 
\pmatrix{0 \cr 1} \},\;  y \in \RR^* \right\} ~\bigcup~
\left\{ \pmatrix{\pm 1 \cr  \pm 1} \right\} ~\bigcup~ 
\left\{ \pmatrix{0 \cr 0 } \right\} ~,
\label{G-orbits1}
\ee
where $\RR^* = \RR \setminus \{ 0 \}$.
The last five points represent Lebesgue-null sets; and hence the
set of representations arising from these orbits will have
Plancherel measure zero (see below). We therefore 
drop them from further discussion. (It ought to be pointed out, 
however, that the first four of these orbits correspond, physically,
to zero-mass systems. Thus, while they do not play any role in 
the Plancherel theory, they are by no means physically negligible.)
On any of the remaining orbits, ${\mathcal O}_{v,y} = 
    y {\rm SO}_{0}(1,1)v$,
the Lorentz group operates
freely. Hence we obtain the parametrization 
\[ \RR \ni \theta \mapsto k  = 
\pmatrix{k_0 \cr \mathbf k} := y \;\Lambda_{\theta}\;v \in {\mathcal O}_{v,y}
~~, \] 
of ${\mathcal O}_{v,y}$, and the measure $d\theta$
is the image of the Haar measure of ${\rm SO}_{0}(1,1)$, 
under this parametrization.
Hence, up to a null set, $\widehat{\RR}^2$ is parametrized by
\be 
\RR \times \{\pmatrix{1 \cr 0},\; \pmatrix{0 \cr 1}\} 
\times \RR^*  \ni (\theta,v,y)
\mapsto y \; \left( \begin{array}{cc}
\cosh \theta & \sinh \theta \\ \sinh \theta & \cosh \theta
\end{array} \right)\; v ~~, \label{dualpar}
\label{G-orbits2}
\ee 
where $\theta$ parametrizes ${\mathcal O}_{v,y}$, and $(v,y)$ 
parametrizes the orbit space. 

 By Mackey's theory of induced representations \cite{Ma68,Ma2}, 
each ${\mathcal O}_{v,y}$ contributes exactly one 
representation class 
$\sigma_{v,y} \in \widehat{\p}$. Denoting the corresponding induced 
representation in this class by $U_{v,y}$, 
its action on  ${\rm L}^2({\mathcal O}_{v,y},d\theta)$ is given by
\[ (U_{v,y}(x,h) f) (k) = e^{i \langle k\;,\; x \rangle}
f(h^{-1}k) ~~,\]
$\langle , \rangle$ denoting the 
dual pairing between $\RR^2$ and $\widehat{\RR}^2$, which we take
(following the physicists' convention) as $\langle k\;,\; x \rangle
= k_0 x_0 - \mathbf k \mathbf x$. Note that this choice of dual
pairing, as opposed to the more conventional
mathematician's choice,  $\langle k\;,\; x \rangle
= k_0 x_0 + \mathbf k \mathbf x$,
does not change the dual action of ${\rm SO}_0(1,1)$.)

 Hence the Plancherel measure $\nu_P$, of the semidirect product 
group $\p = \RR^2 \rtimes {\rm SO}_{0}(1,1)$, 
can be viewed as a measure
on the orbit space $\widehat{\RR^2}/{\rm SO}_{0}(1,1)$, or,
equivalently, on $\{ \pmatrix{1 \cr 0},\; \pmatrix{0\cr 1} \} 
\times \RR^*$.
It is obtained by decomposing the Lebesgue measure of 
$\widehat{\RR}^2$ along the
orbits; in other words, we are looking for a measure 
$\overline{\lambda}$
on $\{ \pmatrix{1\cr 0}, \; \pmatrix{0\cr 1} \} \times \RR^*$ 
such that in the
parametrization (\ref{dualpar}) of $\widehat{\RR}^2$, the Lebesgue
measure is given by $d\theta d\overline{\lambda}(v,y)$.
By computing the Jacobian of (\ref{dualpar}), we obtain
\be
 d\nu_{P}(\sigma_{v,y}) =  
d\overline{\lambda}(v,y) = dv\;y\; dy ~~,
\label{poinplanmeas}
\ee
where $dv$ is just the counting measure on the two-element set
$\{ \pmatrix{1\cr 0}, \; \pmatrix{ 0 \cr 1} \}$.

That we have indeed computed the Plancherel measure
is due to \cite[Theorem 3.3]{FuMa}. ( An alternative
argument could be derived from \cite[II, Theorem 2.3]{KlLi}, or rather,
the proof of that result.) Generally, the procedure for the computation 
of the Plancherel measure of semidirect products $\RR^k \semdir
H$ following Kleppner and Lipsman \cite{KlLi}, involves three steps, which
in this (unimodular) setting may be roughly sketched as follows:
First compute invariant measures on the orbits
(in our case, this was the measure $d\theta$). 
Then compute a unique measure on the orbit space (our 
$\nu_P$) such that first integrating along the
orbits and then integrating over the orbit space gives Lebesgue
measure on the dual. Finally the Plancherel measures
of the little fixed group and the measure on the orbit space
can be combined to give the Plancherel measure of the semidirect
product. In our case, the little fixed groups are trivial,
and in this case the last step reduces to a -- still somewhat
subtle -- normalization issue. (This is discussed at length
in \cite{FuMa}.)

The role played by the decomposition of Lebesgue-measure
for the construction of the Plancherel measure also justifies 
dropping the five orbits from our discussion: They constitute
a Lebesgue-null set, hence the representations arising from the orbits
are a null set with respect to the Plancherel measure.



    The Poincar\'e group above also provides us with an 
easy example of the use 
 of Theorem \ref{th-admissvect}. Since the individual points 
$\sigma_{v,y} \in 
 \widehat{\p}$ have Plancherel measure zero, none of the 
(irreducible) representations 
 $U_{v,y}$ is by itself square-integrable and hence does 
not have admissible 
 vectors. However, it is known \cite{AAG-book,AlAnGa2} 
that if one works on the homogeneous 
 space $\p /T$, where  $T = \{(x_0 , \mathbf 0 ) 
 \; \vert\; x_0 \in \mathbb R \}$ is the subgroup of time
translations, it is possible to 
again obtain admissible vectors
 for these irreducible representations. 
 On the other hand, it should also be possible, 
 according to Theorem \ref{th-admissvect}, to take sets 
of these  representations,
 of finite Plancherel measure such that the corresponding 
(reducible) direct integral 
 representations possess admissible vectors. Such a 
construction was done in 
 \cite{KlSt} (without, however,  relating it to the Plancherel 
theorem). Let us briefly 
 work out the construction again, in the light of 
Theorem \ref{th-admissvect}.
 
    Let $v = (1,0)$ and $\Sigma$ be any Borel subset of 
${\mathbb R}_*$ for which 
$$  0 < \nu_{P} (\Sigma ) = \int_{\Sigma} y\; dy < \infty . $$
Consider the direct integral Hilbert space and the direct 
integral representation on it, 
$$ {\mathfrak H}_{\Sigma} = 
\int_{\Sigma}^{\oplus} {\rm L}^2({\mathcal O}_{v,y},d\theta)\; 
y\;dy, \qquad 
             U_{\Sigma} (x , h ) =  
\int_{\Sigma}^{\oplus} U_{v,y} (x , h ) 
\; y\;dy . $$
Elements $\phi \in {\mathfrak H}_{\Sigma}$ are fields of 
vectors $\phi_{v,y} \in 
{\rm L}^2({\mathcal O}_{v,y},d\theta), \;\; y \in \Sigma$, 
representable 
by functions on ${\mathbb R}^2$ of the type, 
$$ \phi_{v,y} (k) = \phi_{v,y} (k_0 , \mathbf k ) = 
\phi_{v,y}(y\cosh\theta , y\sinh\theta ), \qquad y = 
        \frac {k_0}{\vert k_0\vert}\; 
\sqrt{k_0^2 - {\mathbf k}^2 } \in 
           \Sigma . $$
Explicitly, the representations $U_{v,y} (x , h )$ 
act on the Hilbert spaces 
${\rm L}^2({\mathcal O}_{v,y},d\theta)$ in the manner, 
\begin{eqnarray}
 (U_{v,y} (x , h )\phi_{v,y} )(y\cosh\theta , y\sinh\theta ) 
& = &
        \exp{[iy(x_0 \cosh\theta - \mathbf x \sinh\theta )]} 
\nonumber\\
        &\times & \phi_{v,y} (y\cosh (\theta
          - \xi ) , \;  y\sinh (\theta - \xi ) ),  
\nonumber
\end{eqnarray}
where we have written 
\begin{eqnarray} 
  h  = \Lambda_{\xi} & = & \pmatrix{\cosh\xi  & 
\sinh\xi \cr \sinh\xi & 
\cosh\xi} , \qquad x = \pmatrix{x_0 \cr 
                \mathbf x }, \nonumber \\
  <k \; , \; x > & = & k_0 x_0 - \mathbf k \mathbf x
                   = x_0 \cosh\theta - \mathbf x \sinh\theta 
\nonumber. 
\end{eqnarray}
If we use the variables $(k_0 , \mathbf k )$ rather than $(y , 
\theta )$ to designate points in the orbits, then 
$$ y\; dy\; d\theta  = dk_0 \; d\mathbf k , $$
and 
\be
  (U_{v,y} (x , h )\phi_{v,y} )(k) = \exp [i(k_0 x_0 - \mathbf k 
\mathbf x) ]\; \phi_{v,y} (h^{-1}k ), 
      \qquad y = \frac {k_0}{\vert k_0\vert}\; \sqrt{k_0^2 - 
{\mathbf k}^2 }.  
\label{poincreps}
\ee
      
      Let $\eta = \{\eta_{v,y}\;\vert \; y \in \Sigma\} \in  
{\mathfrak H}_{\Sigma}$ be a 
vector such that $\Vert\eta_{v,y}\Vert = \frac 1{2\pi}$, for 
almost all $y$. Then 
\be
  \langle U_{\Sigma}(x, h) \eta \vert \phi \rangle 
= \int_{\Sigma \times \mathbb R} 
           e^{-i(k_0 x_0 - \mathbf k \mathbf x)} \; 
\overline{\eta_{v,y} (h^{-1}k )}\;
           \phi_{v,y} (k ) \; dk_0 \; d\mathbf k , 
\label{orthogreln}
\ee
and a straightforward computation shows that 
$$ \int_{G} \vert \langle U_{\Sigma} (x , h ) 
\eta \vert \phi \rangle \vert^2 \; d\mu (x , h ) =
   \Vert\phi\Vert^2 ,  \qquad \phi \in {\mathfrak H}_{\Sigma}. $$
Thus, the vector $\eta$ is admissible for the representation 
$ U_{\Sigma}(x, h)$, and 
defining coherent states, $\eta_{(x, h)} = 
U_{\Sigma} (x , h )\eta $, we get the resolution
of the identity on ${\mathfrak H}_{\Sigma}$,
\begin{equation}
   \int_{G}\vert \eta_{(x, h)} \rangle\langle 
\eta_{(x, h)}\vert \; d\mu (x , h ) = {\rm Id}_{\Sigma} . 
\end{equation}

   Before leaving this section, it is worthwhile looking also at the 
affine Poincar\'e group, which is the Poincar\'e group $\p$ just 
considered, together with dilations. Writing this group as 
$\affp = 
{\RR}^2 \rtimes H$, where $H$ now consists of matrices of the 
type
$$ a\;\Lambda_{\theta} = \pmatrix{a\cosh\theta & a\sinh\theta \cr
                           a \sinh\theta & a \cosh\theta }, \quad a > 0, $$
we see that the orbits of $H$ in $\widehat{\RR}^2$ 
consist of the four open cones, 
\be
  {\mathcal C}^{\uparrow , \;\downarrow}_{\pm} = 
H v^{\uparrow ,\;\downarrow}_{\pm} , 
 \qquad  v^{\uparrow}_{\pm} =\pmatrix{\pm 1 \cr 0}, \qquad 
  v^{\downarrow}_{\pm}=\pmatrix{0 \cr \pm 1 },  
\label{G-afforbits1}
\ee
the four semi-infinite lines, 
\be
  {\ell}^\pm_\pm = Hv^\pm_\pm , \qquad v^\pm_\pm = 
\pmatrix{\pm 1 \cr \pm 1},
\label{G-afforbits2} 
\ee
and the singleton consisting of the origin. 
The first four are open free orbits, 
which are unions of orbits of 
the Poincar\'e group $\p$ (see (\ref{G-orbits1})-(\ref{G-orbits2}). 
For example, 
\be  
{\mathcal C}^{\uparrow}_{+} = \bigcup_{y > 0}{\mathcal O}_{v,y}, 
\;\;  v =\pmatrix{1 \cr 0}, 
\qquad 
{\mathcal C}^{\downarrow}_{-} = \bigcup_{y < 0}{\mathcal O}_{v,y}, 
\;\;  v =\pmatrix{0 \cr 1}, 
\label{G-orbits3}
\ee
etc. The remaining five orbits of $\affp$ coincide with the 
five orbits of $\p$ 
which have Plancherel measure zero. The Plancherel 
measure of $\affp$ is just the 
counting measure on the first four orbits, 
the last five orbits again having 
Plancherel measure zero. The unitary irreducible 
representations corresponding to the 
orbits ${\mathcal C}^{\uparrow , \;\downarrow}_{\pm}$ 
are again induced representations 
(from the subgroup of H consisting 
of the identity element alone) and are 
square-integrable. However, the group $\affp$ is nonunimodular 
and hence not 
every vector in these representations is 
admissible (see, e.g., \cite{AAG-book,BeTa}).

\section{Wigner functions}\label{sec-wigfcn}

   Wigner functions are a class of transforms associated 
to elements of the direct integral Hilbert space 
appearing in (\ref{eq:planchtrans}). We denoted this space 
by ${\mathcal B}^\oplus_2$ in Section \ref{sec-planchmeas}. 
The Wigner map associates 
its elements isometrically to square-integrable functions 
on the dual of the Lie algebra of $G$. This dual space foliates 
into orbits under the coadjoint action of the group, the 
invariant components being often identifiable with phase spaces of 
physical systems. 
Motivated by the properties of such a function, 
originally introduced in 
the context of quantum statistical mechanics
by Wigner \cite{Wig}, a general procedure for constructing 
analogous maps 
(applicable to a class of groups admitting square integrable 
representations) was introduced in \cite{AlAtChWo} and further 
discussed in a specific context in \cite{AlKrMu}. Here we 
extend the definition of  a
Wigner function given in \cite{AlAtChWo} to 
representations which are not necessarily square 
integrable, using the Plancherel transform. This will also bring 
into focus the fact that the Wigner 
function, like the wavelet transform, owes its existence 
to the Plancherel transform. 

   It will first be necessary to set out a few details about 
Lie groups and their duals. Again, let $G$ be a Lie 
group with a Type-I regular representation, $\bfrakg$ its Lie 
algebra and $\bfrakg^*$ the dual space of $\bfrakg$. We make the assumption 
that the range of the exponential map, $\bfrakg \ni X \mapsto e^X \in G$, 
is a dense set in $G$, and such that its complement has Haar 
measure zero.
By an {\bf exponential group} we mean a simply connected, connected
solvable Lie group for which the exponential map is a homeomorphism.
A {\bf nilpotent group} is understood to be a simply connected, connected
nilpotent Lie group. In particular, nilpotent groups are exponential.
A Lie group has a natural action on its Lie 
algebra, the adjoint action, $X \mapsto \Ad{x_0}X, \;\; x_0 \in G$, 
defined by, $x_0^{-1}e^X x_0 = e^{[\Ad{x_0}X]}$. The dual of this 
map, acting on $\bfrakg^*$, defines the coadjoint action, 
$\bfrakg^* \ni X^* \mapsto \coAd{x_0}X^*, \;\; x_0 \in G$, via 
$\langle \coAd{x_0}X^*\; ; \; X\rangle = \langle X^* \; ; \; 
\Ad{x_0^{-1}}X \rangle$, where $\langle \; ; \; \rangle$ denotes 
the dual pairing between $\bfrakg$ and $\bfrakg^*$. Orbits of 
vectors in $\bfrakg^*$ under the coadjoint action are the 
{\em coadjoint orbits} of the group $G$. The 
corresponding orbit space, denoted ${\mathcal O}(G)$, has 
a natural quotient topology, and according to the 
Kirillov theory \cite{Ki94,Ki76} for nilpotent groups, 
later extended to exponential groups \cite{LeLu},
this space is homeomorphic to the unitary dual, $\widehat{G}$,
of the group, via the so-called {\em Kirillov map}.
One example is the Poincar\'e group $\p$. For this group, each
coadjoint orbit can be naturally identified with the  
cotangent bundle of a corresponding dual orbit.
More generally speaking, it is known that coadjoint orbits have the 
structure of symplectic manifolds and carry natural invariant 
measures under the coadjoint action, making them resemble 
physical phase spaces. The collection of coadjoint 
orbits exhausts $\bfrakg^*$, allowing for a foliation of the 
type 
$$ \bfrakg^* = \bigcup_{\lambda \in J}{\mathcal O}_\lambda, $$
where ${\mathcal O}_\lambda \in {\mathcal O}(G)$ 
denotes an orbit, parametrized by 
an index (or collection of indices) $\lambda$, and $J$ is the 
corresponding index set. We make the assumption that
the orbit space is a countably separated Borel space, in which case
the Lebesgue measure on $\bfrakg^*$ can be decomposed along these 
orbits, i.e., if 
$dX^*$ denotes this Lebesgue measure, then it is possible to 
write,
\be
  dX^* = \sigma_{\lambda}(X^*_{\lambda} )\;d\kappa (\lambda )\; 
     d\Omega_{\lambda}(X^*_\lambda ), \qquad X^*_\lambda 
  \in {\mathcal O}_\lambda , 
\label{measdisintegr}
\ee
where $\sigma_{\lambda}$ is a positive density defined on 
the orbit ${\mathcal O}_\lambda$ and $d\Omega_\lambda$ the 
(coad)-invariant measure on ${\mathcal O}_\lambda$.  
Note that the assumption on the coadjoint orbits entails
that also the dual space $\widehat{G}$ is a countably
separated Borel space, which is equivalent to the Type-I property
of $G$ (and thus of $\lambda_G$).

The measure $\kappa$ on the orbit space could be continuous
or discrete; whenever it has an atom, it is in fact supported
on finitely many of them. It is only necessary to 
assume that the above disintegration holds
on an open dense set of $\bfrakg^*$, such that its complement 
has Lebesgue measure zero. (Such a decomposition, which is  
sort of a regularity condition, certainly  holds for 
nilpotent groups \cite{ArCo} and semi-direct product groups 
admitting open free orbits \cite{KrAl}, and in these cases, 
the measure $\kappa$ is 
essentially the Plancherel measure.) 

  For each orbit ${\mathcal O}_\lambda$, consider the Hilbert 
space ${\rm L}^2 ({\mathcal O}_\lambda , d\Omega_\lambda )$ and denote 
by $\h^\sharp$ the direct integral Hilbert space, 
$$ \h^\sharp = \int_{J}^{\oplus}{\rm L}^2 ({\mathcal O}_\lambda , 
d\Omega_\lambda )
   \;d\kappa (\lambda ) \simeq {\rm L}^2 (\bfrakg^*) ,$$
where we used the measure disintegration (\ref{measdisintegr}) to 
canonically identify ${\rm L}^2 (\bfrakg^*)$ with the direct integral.
Elements in $\h^\sharp$ are fields of vectors, $\Phi = 
\{\Phi_\lambda \in {\rm L}^2 ({\mathcal O}_\lambda , d\Omega_\lambda )
\}_{\lambda \in J}$ with the norm,
$$ \Vert\Phi\Vert^2 = \int_J \Vert\Phi_\lambda\Vert^2\; 
   d\kappa (\lambda ), $$
the norm inside the integral being taken in 
${\rm L}^2 ({\mathcal O}_\lambda , d\Omega_\lambda )$. If the measure 
$\kappa$ is discrete, then clearly the 
integral would just be a sum. 
The Wigner map will be defined as a linear isometry, 
$W: {\mathcal B}^\oplus_2 \longrightarrow  \h^\sharp$. Let $N_0 
\in \bfrakg$ be the maximal symmetric set (i.e., $N_0$ includes 
the origin and $X \in \bfrakg
\Rightarrow -X \in \bfrakg$) such that its image under the 
exponential map is dense in $G$ and such that the complement of
this image set has Haar measure zero. For any $f \in {\rm L}^2 (G), \;
f(e^X )$ defines a function on $N_0$. We transfer the left Haar 
measure $\mu_G$ to $N_0$, using the exponential map and write, 
\be
  d\mu_G (g) = d\mu_G (e^X ) = m(X)\; dX, \qquad X \in N_0 , 
\label{eq:liealgmeas}
\ee
where $dX$ is the Lebesgue measure of $\bfrakg$ and $m$ an 
appropriate, positive density function. It is not hard to see 
that 
\be 
   m(X) = \vert \mbox{\rm det}\;[-F(\mbox{\rm ad} X)]\vert , 
\label{mfcn}
\ee
where $F$ is the function (\ref{F-fcn}) defined in the Appendix 
and $\mbox{\rm ad} X$  the linear transformation on $\bfrakg$
defined by $\mbox{\rm ad} X (Y) = [X, Y], \;\; Y \in \bfrakg$.

   Let us next define a {\em modified} Fourier transform, 
${\mathcal F}: {\rm L}^2(G) \longrightarrow \h^\sharp$ as 
\be
  ({\mathcal F}f)_\lambda (X^*_\lambda ) = 
  \frac {[\sigma_\lambda (X^*_\lambda )]^{\frac 12}}
         {(2\pi )^{\frac n2}}\;\int_{N_0} e^{-i\langle X^*_\lambda ;
       X\rangle} f(e^X ) [m(X)]^{\frac 12}\; dX ,
\label{eq:modFourtrans}
\ee
at least on ${\rm L}^1(G) \cap {\rm L}^2(G)$, and extend by
continuity.
(Note, we are assuming the dimension of the group $G$, 
and hence of its 
Lie algebra, to be $n$). 
Since the complement of the set $N_0$ is of (Haar) measure zero,
this map is easily seen to be an isometry. 
For nilpotent groups (\ref{eq:modFourtrans}) simplifies considerably
since all involved density functions (i.e., $\sigma_{\lambda}$, $m$) are
identical one.

\begin{Defn}
The composite transformation 
\be
  {\mathfrak W} : = {\mathcal F}\circ {\mathcal P}^{-1}: 
    {\mathcal B}^\oplus_2 \longrightarrow \h^\sharp , 
\label{eq:Wigmap}
\ee
where $\mathcal P$ is the Plancherel transform in
(\ref{eq:planchtrans}), is 
called the {\bf Wigner map} and for any 
$A \in {\mathcal B}^\oplus_2$, the function 
\be
  W(A \vert X^*_\lambda) := 
   ({\mathfrak W}A)_\lambda (X^*_\lambda), \qquad 
   X^*_\lambda \in {\mathcal O}_\lambda  
\label{eq:Wigfcn}
\ee
is called the {\bf Wigner function} of $A$, restricted to the orbit 
${\mathcal O}_\lambda$. 
\end{Defn}

  For any $A \in \h^\sharp$ which satisfies the conditions of 
Theorem \ref{Pl-Inv},  using (\ref{eq:invexp}) we obtain the 
following explicit expression for its Wigner function: 
\bea
   W(A\vert X^*_\lambda ) & = &
    \frac {[\sigma_\lambda (X^*_\lambda )]^{\frac 12}}
         {(2\pi )^{\frac n2}}\;\int_{N_0} e^{-i\langle X^*_\lambda ;
       X\rangle} \nonumber \\ 
   & \times & \left[\int_{\widehat{G}}\mbox{\rm tr}\;
       (U_{\sigma}(e^{-X} )[A(\sigma )C_\sigma^{-1}])\;
       [m(X)]^{\frac 12}\;d\nu_G (\sigma )\right]\; dX ,
\label{eq:explWigfcn}
\eea
provided the inverse Plancherel transform of $A$ is integrable.
The inverse of this transform can be computed using (\ref{eq:invexp}) and 
standard Fourier transform methods. We get, 
\bea
   A(\sigma) & = & \frac 1{(2\pi )^{\frac n2}}
\Bigg[\;\Big\{\int_{N_0}\Big[\int_J 
     \int_{{\mathcal O}_\lambda} e^{i\langle 
X^*_\lambda \; ;\;  X\rangle}\;
      W(A\vert X^*_\lambda ) \; U_\sigma (e^X )\nonumber \\
    &\times & [\sigma_\lambda (X^*_\lambda )\; 
m(X)]^{\frac 12}\;d\Omega_\lambda\;
      d\kappa (\lambda )\Big]\; dX\Big\}\; C_\sigma^{-1}\Bigg] ,
\label{invexplWigfcn}
\eea
the extreme pair of square brackets implies taking the 
closure of the operator
involved. 

  A few properties of the Wigner map can easily be established 
from its definition. We collect these into the theorem below. 
The proof 
involves straightforward computations, similar to those 
done to obtain analogous results in \cite{AlAtChWo}, and we omit it. 

   On ${\mathcal B}_2^\oplus$ and $\h^\sharp$ we define the two 
unitary 
representations, $U^\oplus$ and $U^\sharp$,  of the group $G$:
\be
  (U^\oplus (x) A)(\sigma ) = U_\sigma (x) A(\sigma) 
U_\sigma^* (x), \qquad x \in G, 
\label{eq:uplus}
\ee
the above relations holding for almost all $\sigma \in 
\widehat{G}$ (w.r.t. $\nu_G$), 
and
\be
  (U^\sharp (x)\Phi )_\lambda (X^*_\lambda ) = 
    \Phi_\lambda (\coAd{x^{-1}} X^*_\lambda ), \qquad x \in G , 
\label{eq:usharp}
\ee
holding for almost all $\lambda \in J$ (w.r.t. $\kappa$). 

\begin{thm} \label{thm:wigner_map}
   The Wigner map is a linear isometry, which intertwines 
the representation 
$U^\oplus$ of $G$ with the representation $U^\sharp$. The 
corresponding Wig\-ner function
satisfies the {\bf overlap condition}, 
\be
  \int_J \left[\int_{{\mathcal O}_\lambda}
   \overline{W(A^1 \vert X^*_\lambda )}\; W(A^2 \vert X^*_\lambda )
    \; d\Omega_\lambda \right] \; d\kappa (\lambda ) =
    \langle A^1 \vert A^2 \rangle_{{\mathcal B}^\oplus_2}
     , \qquad A^1, A^2 \in 
   {\mathcal B}^\oplus_2 , 
\label{eq:overlap}
\ee
and the {\bf covariance condition},
\be
  W(U^\oplus (x)A \vert X^*_\lambda ) = 
   W(A \vert \coAd{x^{-1}}X^*_\lambda ),
\label{eq:covcond}
\ee
for all $x \in G$, and almost all $X^*_\lambda \in 
{\mathcal O}_\lambda$ (w.r.t. $\Omega_\lambda$). If 
$A = A^*$ is self adjoint, its Wigner function 
is real, i.e., 
\be
  \overline{W(A \vert X^*_\lambda )} = W(A \vert X^*_\lambda ),  
\label{eq:reality}
\ee
almost everywhere. 
\end{thm}

   Note that if $A^1 , A^2 \in {\mathcal B}_2^\oplus$ satisfy the 
conditions of Theorem \ref{Pl-Inv}, then Theorem \ref{Pl-Thm} 
together with (\ref{eq:invexp}) implies the {\bf orthogonality
relation}
\bea
\lefteqn{
 \int_G\left\{\int_{\widehat{G}} \overline{{\rm tr}([ 
 U_{\sigma}(x)^*A^1 (\sigma)C_{\sigma}^{-1}])} 
  d\nu_G(\sigma)\right. }\nonumber\\ 
& &\times\left.\int_{\widehat{G}} 
{\rm tr}([ U_{\sigma'}(x)^*A^2 (\sigma')C_{\sigma'}^{-1}]) 
  d\nu_G(\sigma')\right\}\;d\mu (x) 
  = \langle A^1\; \vert \; A^2\rangle_{{\mathcal B}_2^\oplus},
\label{Pl-orthog}
\eea
which is equivalent to the overlap condition (\ref{eq:overlap}). 
If now $\sigma \in \widehat{G}$ has positive Plancherel measure, 
then the above relation implies the restricted orthogonality 
relation, 
\begin{eqnarray*}
 \lefteqn{\hspace{-85pt}\int_G 
 \overline{{\rm tr}([U_{\sigma}(x)^*A^1 
   (\sigma)C_{\sigma}^{-1}])}\; 
{\rm tr}([ U_{\sigma}(x)^*A^2 (\sigma)C_{\sigma}^{-1}]) 
  \;d\mu (x)} \\ 
  & & = \mbox{\rm tr}\; [A^1(\sigma )^*\;A^2 (\sigma )],
\end{eqnarray*}
familiar from the theory of square integrable group representations
\cite{AAG-book,GrMoPa}. This equation was the basis for the 
construction of Wigner functions, for square integrable 
group representations,  in \cite{AlAtChWo}.

\begin{rems}
A few comments are in order here:

\medskip
\begin{enumerate}

\item[(a)] Generally, the range of the Wigner map $\mathfrak W$ 
is a closed, proper 
subspace of $\h^\sharp$, which we denote by $\h^\sharp_W$. 
If we restrict 
the Wigner map to a subspace of ${\mathcal B}^\oplus_2$ of the type 
$$ {\mathcal B}^\Sigma_2  = \int_\Sigma^\oplus {\mathcal B}_2 
(\h_\sigma )
   \;d\nu_G (\sigma) , $$
where $\Sigma$ is subset a of $\widehat{G}$, 
such that $\nu_G (\Sigma ) \neq 0$, 
then clearly its range ${\mathfrak W} ({\mathcal B}^\Sigma_2 )$ 
is a closed subspace of 
$\h^\sharp_W$. In this case, the integral over $\widehat{G}$ 
in (\ref{eq:explWigfcn}) has
to be replaced by an integral over $\Sigma$, however 
the expressions (\ref{invexplWigfcn})
and (\ref{eq:overlap}) remain unchanged. In particular, 
if $\Sigma$ is a discrete subset,
the representations $\sigma \in \Sigma$ are 
square-integrable and we recover the 
results of \cite{AlAtChWo}. 

\medskip

\item[(b)] Suppose that the group $G$ is exponential and 
 assume, moreover,  that it is
 a Type-I group. Note that the homeomorphism property of the 
 Kirillov map entails that the coadjoint orbit space is
 a countably separated Borel space; in particular the measure 
disintegration (\ref{measdisintegr}) exists.

 Then we have on the one hand a mapping
 \[ W : \int^{\oplus}_{\widehat{G}} {\mathcal B}_2({\mathfrak H}_{\sigma}) 
  d\nu_G(\sigma) \to  \int_{J}^{\oplus}{\rm L}^2 ({\mathcal O}_\lambda , 
d\Omega_\lambda )
   \;d\kappa (\lambda ) \simeq {\rm L}^2 (\bfrakg^*) ,\]
 between the direct integral spaces, and on the other hand the inverse
 of the Kirillov map, which gives rise to a bijection $\widehat{G} \to J$,
 $\sigma \mapsto \lambda_\sigma$. It is thus a natural question to 
 ask whether the Wigner map is decomposable, i.e.,
 if there exists a field of operators 
 \[ W_{\sigma} :  {\mathcal B}_2({\mathfrak H}_{\sigma}) \to 
 {\rm L}^2 ({\mathcal O}_{\lambda_\sigma} , d\Omega_{\lambda_{\sigma}} )
 \]
 such that for almost all $\sigma \in \widehat{G}$ and
 almost all $X_{\lambda_{\sigma}}^* 
\in {\mathcal O}_{\lambda_{\sigma}}$,
 \[ W(A| X_{\lambda_{\sigma}}^*) = \left[ W_{\sigma} A (\sigma)\right] 
 (X_{\lambda_{\sigma}}^*) ~~.\]
 The existence of such a field of operators is not just of mathematical
 interest, but also desirable from a physical point of view: 
 The coadjoint orbits have a natural interpretation as phase
 spaces of physical systems, but the dual $\bfrakg^*$, as a disjoint union of 
such phase spaces,
 does not usually have a natural interpretation, except in some cases, where one 
might look upon a set of orbits as constituting the phase space of a composite 
physical system. Correspondingly, the space 
 ${\rm L}^2 ({\mathcal O}_{\lambda_\sigma}, d\Omega_{\lambda_{\sigma}} )$
 has a simpler interpretation than ${\rm L}^2(\bfrakg^*)$. A similar reasoning
 applies to the representations.

 A related question concerns the supports of the Wigner functions.
Even when the Wigner map is restricted to a subspace such 
as ${\mathcal B}^\Sigma_2$ as in part (a), the corresponding Wigner 
functions could in general have supports on orbits which are 
not associated to the representations in $\Sigma$
(see example of the Poincar\'e group below). This 
is possible even for representations which arise from 
semidirect product groups admitting open free orbits \cite{KrAl}.
It is obvious that whenever the Wigner map is decomposable,
the supports of the Wigner functions of elements in
${\mathcal B}_2^\Sigma$ are contained (up to a null set)
in the coadjoint orbits corresponding to $\Sigma$; we expect
the converse of this statement to hold as well.

It turns out that these questions have been addressed, and to a large 
extent solved, in the context of star products: First of all,
the nilpotent Lie groups for which
the Wigner transform is decomposable are precisely those
for which almost all coadjoint orbits are affine subspaces
\cite{Lu}. If a nilpotent Lie group does not fulfill this
condition, the modified Fourier transform (\ref{eq:modFourtrans}) 
can be replaced by an {\bf adapted Fourier transform}. 
Following \cite{ArCo,ArGu}, the adaptation consists in 
constructing a suitable mapping $\alpha :
{\mathfrak g} \times {\cal V} \to \RR$, polynomial in the
elements of ${\mathfrak g}$ and rational in the elements of a
suitably chosen open conull subset ${\cal V} \subset
{\mathfrak g}^*$. The specific construction of $\alpha$ first ensures that
defining
\begin{equation} \label{eq:adFourtrans}
 {\cal F}_{ad} (f) (X^*) = \frac {1}
         {(2\pi )^{\frac n2}}\;\int_{{\mathfrak g}} e^{-i \alpha(X,X^*)} 
       f(e^X ) ~dX ~~,
\end{equation}
for $f$ in the Schwartz space of the group, induces a 
a unitary map ${\rm L}^2(G) \to {\rm L}^2({\mathfrak g}^*)$.
Secondly, the {\bf adapted Wigner map}
${\mathfrak W}_{ad} = {\cal F}_{ad} \circ {\cal P}^{-1}$
has all the properties of the Wigner map collected in Theorem 
\ref{thm:wigner_map}, and is in addition decomposable.
Extensions of this construction to certain solvable groups
exist \cite{ArCoLu}. It seems worthwhile to explicitly work out adapted Wigner
transforms for concrete examples. This might also provide additional criteria
for the choice of $\alpha$, which is apparently not unique.
\end{enumerate}
\end{rems}

\section{Some examples}\label{sec-examples} 
Let us go back to the Poincar\'e groups $\p$  and $\affp$, studied in Section 
\ref{sec-expoincare}, and 
explicitly compute the Wigner functions for them. 

\subsection{The Poincar\'e group $\p$}

We start by writing a general element 
of $\p$ as a $3\times 3$ matrix, 
$$ 
  (x , h ) = \pmatrix{h & x \cr {\bf 0}^T & 1}, \qquad {\bf 0}^T 
= (0,0), $$
where $x$ and $h$ are as defined earlier (in Section 
\ref{sec-expoincare}). The Lie 
algebra $\mathfrak p$ is generated by the three elements, 
\be 
 Y^1 = \pmatrix{0 & 1 & 0 \cr 1 & 0 & 0 \cr 0 & 0 & 0}, \quad 
   Y^2 = \pmatrix{0 & 0 & 1 \cr 0 & 0 & 0 \cr 0 & 0 & 0}, \quad
   Y^3 = \pmatrix{0 & 0 & 0 \cr 0 & 0 & 1 \cr 0 & 0 & 0}, 
\label{poincliealg}
\ee
which satisfy the commutation relations,
$$ [Y^1 , Y^ 2 ] = Y^3 , \qquad  [Y^1 , Y^ 3 ] = Y^2 , \qquad 
   [Y^2 , Y^ 3 ] = 0. $$
A general element $X \in \mathfrak p$ can be written as 
(see (\ref{Liematrix}) in the Appendix),
$$
    X = \pmatrix{X_q & x_p \cr {\bf 0}^T & 0}, \quad X_q := 
  X_q (0, \theta ) =
    \pmatrix{0 & \theta \cr \theta & 0}, \;\; \theta \in \RR , \quad 
    x_p = \pmatrix{\xi_1 \cr \xi_2} \in {\RR}^2 , $$
so that 
$$ 
    e^X = \pmatrix{e^{X_q} & F(X_q )x_p \cr {\bf 0}^T & 1}, $$
$F(X_q )$ being the matrix function defined in 
(\ref{matrixfcns}) in the Appendix. Following (\ref{eq:liealgmeas}), 
the Haar measure $d\mu(x, h) = dx\;d\theta$, expressed in terms 
of the Lie algebra variables $x_p , \theta$, becomes
$$ d\mu (x, h ) = m(x_p , \theta )\; dx_p\; d\theta , \qquad 
        dx_p = d\xi_1 \;d\xi_2 , $$
and the density $m(x_p , \theta )$ is easily calculated to be
(see (\ref{matrixfcns}) in the Appendix),
\be
   m(x_p , \theta ) = \mbox{\rm det}\;[F(X_q )] 
   = \sinch^2 (\frac{\theta}2).
\label{poincliealgden}
\ee

   The adjoint action of $\p$ on $\mathfrak p$, given by $X 
\longrightarrow X' = (x, h )X (x, h )^{-1}$,  leads to the 
transformation 
$$  
\pmatrix{x'_p \cr \theta ' } = M (x,h ) 
\pmatrix{x_p \cr \theta  }, \qquad 
    M(x,h) = \pmatrix{h & \sigma_1 x \cr {\bf 0}^T & 1}, $$
of the variables $x_p , \theta $, where $\sigma_1$ is 
the $2\times 2$ matrix 
defined in the Appendix. Let ${\mathfrak p}^*$ denote 
the dual space of 
$\mathfrak p$. We write elements $X^* \in  {\mathfrak p}^*$ 
in terms of the 
dual basis $\{Y_1^* ,  Y_2^* , Y_3^*\}$ as $X^* = 
\gamma Y_1^* + k_0 Y_2^* 
+ \mathbf k Y_3^*$ and compute the coadjoint action, in 
terms of a matrix 
$M^\sharp (x, h)$ acting on the variables $k, \gamma$,
$$
   \pmatrix{k' \cr \gamma'} = M^\sharp (x, h)\pmatrix{k \cr \gamma} =
   M(-h^{-1}x, h^{-1})^T \pmatrix{k \cr \gamma}, 
      \qquad k = \pmatrix{k_0 \cr \mathbf k} \in {\RR}^2 ,  $$
to obtain,
\bea 
  k' & = & h^{-1}k \nonumber \\
  \gamma' & = & \gamma - x^T \sigma_1 k' , \qquad x^T = 
(x_0 , \mathbf x ).
\label{poincoadaction}
\eea
Using these relations, all the coadjoint orbits of $\p$ in 
$\RR^3 \simeq {\mathfrak p}^*$
can now be calculated. Indeed, introducing the vectors $yv$, 
defined 
in (\ref{G-orbits1}), we get the coadjoint orbits 
\be
  {\mathcal O}^*_{v,y} = \left\{ y M^\sharp (x, h)\pmatrix{v \cr 0} 
   \;\Big| \; (x, h) \in \p \right\}, 
\label{poincoadorb}
\ee
which, taken together for all $y, v$, exhaust $\RR^3$. 
It is also clear from 
(\ref{poincoadaction}) that these orbits are precisely the 
{\em cotangent bundles}, 
${\mathcal O}^*_{v,y} = T^*{\mathcal O}_{v,y}$, of the orbits 
${\mathcal O}_{v,y} 
= y SO_0 (1,1) v$ computed in Section (\ref{sec-expoincare}).  

   Explicitly, let us take the set 
\be
  \Sigma = \{yv \;\vert\;
y > 0 \} \subset \widehat{\p}, \qquad  v = \pmatrix{1 \cr 0}.
\label{dualsubset}
\ee

Points in the corresponding 
coadjoint orbits can then be parametrized as
$$ {\mathcal O}^*_{v,y} = \{ (k, \gamma ) \in {\RR}^3 \;\vert \;
       k = (k_0 , {\mathbf k}) , \;\;     
        k_0 > 0, \;\; y = \sqrt{k_0^2 - {\mathbf k}^2} \} , $$
and  the invariant measure under the 
coadjoint action calculated to be 
$$ d\Omega_{v,y} (k, \gamma ) = 
\frac {d{\mathbf k}}{k_0}\;d\gamma . $$
Of course, we could also use the alternative 
coordinatization $(\theta , -\gamma )$
for ${\mathcal O}^*_{v,y}$, where $\theta = 
-\tanh^{-1}({\mathbf k}/k_0 )$, which are 
actually the {\em Darboux coordinates} for this orbit, 
and then the invariant 
measure would simply be $d\theta\; d\gamma$. However it will 
be more useful, for 
the purposes of computing the Wigner function, to use the 
$(k, \gamma )$ 
coordinates.  The Lebesgue measure on ${\mathfrak p}^*$ in the 
$(k, \gamma )$ coordinates is $dk_0\;d{\mathbf k}\; d\gamma$, and 
making the change of variables, $(k_0, {\mathbf k}, \gamma ) 
\longrightarrow (y , {\mathbf k}, \gamma )$ we get the measure 
disintegration along the coadjoint orbits in $\Sigma$ 
(see (\ref{measdisintegr})),
\bea
   dk_0\; d{\mathbf k}\; d\gamma & = & y\;dy \;
   \frac{d\mathbf k}{\sqrt{{\mathbf k}^2 + y^2}}\;d\gamma
      = \sigma_{v,y}(k,\gamma )\;d\kappa (v,y)\;
       d\Omega_{v,y}(k, \gamma ), \nonumber \\
\mbox{\rm with} & \quad &
    \sigma_{v,y}(k,\gamma ) \equiv 1, \qquad d\kappa (v,y) = y\;dy.
\label{poinmeasdis}
\eea
Thus, in this case, the measure $\kappa$ is precisely equal to 
the Plancherel measure $\nu_P$, restricted to $\Sigma$ 
(see (\ref{poinplanmeas})). 

   We are now in a position to explicitly compute the
Wigner map
$$ \mathfrak W : {\mathcal B}^\Sigma_2 \longrightarrow 
         \h^\sharp , $$
for $\p$, restricted to the subset  $\Sigma$ defined in 
(\ref{dualsubset}). Since $\p$ is unimodular, the Duflo-Moore 
operators $C_\sigma$ are trivial and in fact, using relations 
such as (\ref{orthogreln}), it can be seen that for almost
all $yv \in \Sigma$, the corresponding Duflo-Moore operator
is $C_{v,y} = 2\pi\;{\rm Id}_{{\mathfrak H}} $ on the representation
space ${\mathfrak H} = {\rm L}^2 ({\mathcal O}_{v, y} , d\theta )$.  Let us consider
elements in  ${\mathcal B}^\Sigma_2$ which are of the type
$$ A = \{A(v, y)\}_{yv \in \Sigma}, \;\; A(v,y) = 
\vert\phi_{v,y}\rangle\langle\psi_{v,y}\vert, \qquad 
   \phi_{v,y}, \; \psi_{v,y} \in 
    {\rm L}^2 ({\mathcal O}_{v, y} , d\theta ) .$$
A tedious but straightforward manipulation, after inserting the 
various quantities into the expression (\ref{eq:explWigfcn}) for 
the Wigner function and using relations such as (\ref{matrixfcns}),
yields the final expression,
\bea
  W(A \vert k_{v,y} \; , \gamma )& = & \frac 1{(2\pi)^{\frac 12}}\;
      \int_{\mathbb R}e^{i\theta\gamma}\; 
   \overline{\psi_{v, y(\theta )} 
   \Big(\frac {e^{-\frac {\theta\sigma_1}2}\; \sigma_3\;k_{v,y}}
     {\sinch (\frac {\theta}2 )}\Big)}\;
    \frac 1{\sinch (\frac {\theta}2 )}\nonumber\\ 
   & \times & \phi_{v, y(\theta )} 
   \Big(\frac {e^{\frac {\theta\sigma_1}2}\; \sigma_3\; k_{v,y}}
     {\sinch (\frac {\theta}2 )}\Big)\;d\theta , \qquad 
    y(\theta ) = \frac y{\sinch (\frac {\theta}2 )} . 
\label{poinWigfcn}
\eea
Here, $k_{v, y} \in {\mathcal O}_{v, y}$, and the point 
$(k_{v,y}\; , \gamma ) \in T^* {\mathcal O}_{v, y}$ and hence 
the above expression is for the Wigner function 
restricted to the orbit 
$T^* {\mathcal O}_{v, y}$. However, it ought to be noted that 
its value on any orbit receives contributions from 
vectors $\phi_{v, y(\theta )},\; \psi_{v, y(\theta )}$ coming 
from representations associated to all the orbits in $\Sigma$.
Hence we see that the Wigner map is not decomposable.
(Completely the opposite situation is true for the affine Poincar\'e 
group, as will be shown in Theorem \ref{thm:Wigfcnsupp} below.) 

   Using (\ref{poincoadaction}) and (\ref{matrixfcns2}) we 
directly verify the covariance condition (\ref{eq:covcond}),
$$
  W(U^\oplus_\Sigma (x,h)A \vert k_{v,y}\; , \gamma ) = 
    W(A \vert k'_{v,y}\; , \gamma' ), \quad 
  \pmatrix{k'_{v,y} \cr \gamma'} = M^\sharp (x, h)^{-1}
   \pmatrix{k_{v,y}\cr \gamma},  $$
$(x, h ) \in \p$, with 
$$  (U^\oplus_\Sigma (x,h)A)(v,y) = U_{v,y}(x, h)\vert\phi_{v,y}\rangle
          \langle\psi_{v,y}\vert U_{v,y}(x, h)^*, \qquad 
   yv \in \Sigma , $$
and the overlap condition, 
\begin{eqnarray*}
\lefteqn{\hspace{-75pt}\int_{{\mathcal O}_{v,y}\times \RR^+}
  \overline{W(A^1 \vert k_{v,y} \; , \gamma )}
  W(A^2 \vert k_{v,y} \; , \gamma )\;
  d\Omega_{v,y}(k_{v,y} \; , \gamma )\;y\;dy =}\\ 
& &  \int_{\RR^+}\langle\psi^2_{v,y} \vert\psi^1_{v,y}
  \rangle\langle\phi^1_{v,y}\vert\phi^2_{v,y}\rangle\;y\;dy, 
\end{eqnarray*}
where, $A^i = \{\vert\phi^i_{v,y}\rangle
\langle\psi^i_{v,y}\vert\}_{yv \in \Sigma}, \;\; i=1,2$.

   Consider now the open forward {\em light cone} ${\mathcal C}^\uparrow_+$ 
(see (\ref{G-orbits3}),
$$
  {\mathcal C}^\uparrow_+ = 
\left\{\pmatrix{k_0 \cr \mathbf k }\in \RR^2 \;
      \vert\; k_0 > 0, \;\; k_0^2 > {\mathbf k}^2 \right\} .$$ 
Using the coordinates $(y , \theta ) = (\sqrt{k_0^2 - {\mathbf k}^2},
-\tanh^{-1} ({\mathbf k}/{k_0}))$, the invariant measure under the 
action of $SO_0 (1,1)$ is clearly $y\;dy\;d\theta$ and  the 
Hilbert space ${\rm L}^2 ( {\mathcal C}^\uparrow_+ , y\;dy\;d\theta )$ is 
naturally isomorphic to the direct integral Hilbert space 
  $$ \h_\Sigma = \int_{\Sigma}^\oplus {\rm L}^2 ({\mathcal O}_{v,y}\;, 
      d\theta )\; y\;dy .$$
The corresponding direct integral representation $U_\Sigma$ can 
thus be expressed by its action on 
${\rm L}^2 ( {\mathcal C}^\uparrow_+ , y\;dy\;d\theta )$ in the 
manner 
$$ (U_\Sigma (x , h )\phi )(y, \theta ) = \exp [iy (x_0 \cosh\xi 
    -{\mathbf x}\sinh\xi )]\;\phi (y , \theta - \xi ), $$
$\phi \in {\rm L}^2 ( {\mathcal C}^\uparrow_+ , y\;dy\;d\theta )$ and 
$\xi$ being the hyperbolic angle of the transformation $h$. Thus, 
the representation $U_\Sigma$ is  precisely the Fourier transform of 
the quasi-regular 
representation of $\p$, restricted to the Hilbert space
${\rm L}^2({\mathcal C}^\uparrow_+ , \; y\;dy\;d\theta )$, of 
functions with support in the forward light cone. 
The Wigner function (\ref{poinWigfcn}) can now be thought of as a 
function on ${\mathcal C}_+^\uparrow \times \RR \simeq 
\bigcup_{y>0}T^* {\mathcal O}_{v,y}$ and, 
written in terms of the variables $(y, \theta , -\gamma )$ (the 
invariant measure under the coadjoint action being $y\;dy\;d\theta\;
d\gamma$), it becomes 
\bea
  W'(\psi , \phi \; \vert\; \theta, \gamma; \; y) & 
= & \frac 1{(2\pi)^{\frac 12}}\;
      \int_{\mathbb R}e^{-i\xi\gamma}\; 
   \overline{\psi 
   \Big(\frac y{\sinch(\frac {\xi}2 )} , \; \theta - \frac {\xi}2\Big)}\;
    \frac 1{\sinch (\frac {\xi}2 )}\nonumber\\ 
   & \times & \phi \Big(\frac y{\sinch(\frac {\xi}2 )} , \; \theta + \frac 
{\xi}2\Big)
   \;d\xi , \quad \theta, \; \gamma \in \RR, \;\; y> 0, 
\label{poinWigfcn2}
\eea
$\phi , \psi \in {\rm L}^2({\mathcal C}^\uparrow_+ , \; y\;dy\;d\theta )$. 
It ought to be noted, however, that although in this way of 
writing the Wigner function, $W'(\psi , \phi \; \vert\; k, \gamma )$
is sesquilinear in $\psi , \; \phi \in  
 {\rm L}^2({\mathcal C}^\uparrow_+ ,\; y\;dy\; d\theta )$, and can be extended
to the linear span of rank-one operators $\vert\phi\rangle\langle\psi\vert$, 
it cannot be used to define an isometric map between 
${\mathcal B}_2 ({\rm L}^2({\mathcal C}^\uparrow_+ ,\; y\;dy\; d\theta))$ and
${\rm L}^2({\mathcal C}^\uparrow_+ \times \RR ,\; y\;dy\; d\theta\;d\gamma )$, 
since
\bea 
\int_{{\mathcal C}^\uparrow_+ \times \RR }\vert 
W'(\psi , \phi \; \vert\; \theta, \gamma; \; y) \vert^2\; y\;dy\; d\theta d\gamma
  & =  & \int_{\RR^+ \times \RR^2}\vert\psi (y, \theta )\vert^2\;\vert\phi
  (y, \xi )\vert^2\nonumber\\
  & \times &  y\;dy\;d\theta\;d\xi
   \neq  \Vert\phi\Vert^2\;\Vert\psi\Vert^2 \; . \nonumber \eea
Thus $W'(\phi , \psi \;\vert\; \theta , \gamma ; \; y)$ is not a Wigner function for the
operator $\vert\phi\rangle\langle\psi\vert  \in
{\mathcal B}_2 ({\rm L}^2({\mathcal C}^\uparrow_+ ,\; y\;dy\; d\theta))$, in the
sense of our definition (hence the   use of the
altered notation $W'$). On the other hand, physically the
representation $U_\Sigma$ refers to systems of 
relativistic particles of all possible (positive) masses and
$W'(\psi , \psi \; \vert\; y, \theta , \gamma )$ can serve as the Wigner 
function for the state of a system consisting of a cluster
of masses.  Furthermore, this form of the Wigner function is particularly
simple looking and bears a striking resemblance to the original Wigner 
function \cite{Wig} (see (\ref{standWigfcn2}) below). 

\subsection{The affine Poincar\'e group $\affp$}
Elements, $(x, ah) \in \affp , \;\; a > 0 , \; h \in SO_0 (1,1)$ can be 
represented by matrices of the form 
$$ 
  (x, ah ) = \pmatrix{ah & x \cr \mathbf 0 & 1} . $$
The group is nonunimodular, with left and right Haar measures, 
$$ 
  d\mu_{\ell}(a,h, x) = \frac 1{a^3}\; 
     {dx_0 \;d{\mathbf x}\;da\;d\theta}, \qquad 
d\mu_{r}(a,h, x) = \frac 1{a}\; 
     {dx_0 \;d{\mathbf x}\;da\;d\theta}. $$
As discussed at the end of Section \ref{sec-expoincare}, there are 
four irreducible representations of $\affp$, corresponding to the 
four open free orbits ${\mathcal C}_\pm^{\uparrow , \;\downarrow}$ 
(see (\ref{G-afforbits1})) which 
are square integrable, and these are 
the only ones which contribute to the Plancherel measure. It will 
be enough to work out the Wigner function for the one orbit 
${\mathcal C}_+^{\uparrow}$, for the other three are entirely 
similar. The Hilbert space of the irreducible representation 
$U_+^\uparrow$, associated to this orbit,  is 
${\rm L}^2( {\mathcal C}_+^{\uparrow}, dk_0 \; d{\mathbf k})$ and 
$$
  (U_+^\uparrow (x, ah )\phi )(k ) = ae^{i\langle k \; , x\rangle}
                  \phi (ah^{-1}k). $$
The Duflo-Moore operator $C$ for this representation is unbounded, 
acting on ${\rm L}^2( {\mathcal C}_+^{\uparrow}, dk_0 \; d{\mathbf k})$  
in the manner (see \cite{AAG-book, BeTa}), 
$$ 
 (C\phi )(k) = \frac {2\pi}{\vert k_0^2 - 
    {\mathbf k}^2\vert^{\frac 12}}\;\phi (k). $$
Recall that the orbit ${\mathcal C}_+^{\uparrow}$ is characterized 
by $k_0 > 0 , \;\; k_0^2 > {\mathbf k}^2$ and the invariant measure 
on it under the action of the group elements $ah$ is 
$\frac {dk_0 \; d{\mathbf k}}{k_0^2 - {\mathbf k}^2}$.

   The Lie algebra ${\mathfrak p}_{\rm Aff}$ is four dimensional,
being generated by the three elements (\ref{poincliealg}) of the
Lie algebra of $\p$ together with ${\mathbb I}_3$, the $3\times 3$
identity matrix. Computation of the coadjoint orbits is routine. The 
one which concerns us here is the cotangent bundle, 
$T^*{\mathcal C}_+^{\uparrow}$, of the orbit 
${\mathcal C}_+^{\uparrow}$. Denoting its elements by 
$(\gamma , k ), \;\; \gamma = \pmatrix{\gamma_1 \cr \gamma_2} \in 
{\mathbb R}^2 , \;\; k \in {\mathcal C}_+^{\uparrow}$, 
using relations such as (\ref{matrixfcns2}) - (\ref{matrixfcns3})
the coadjoint action is computed to be, 
\bea
    k \longrightarrow k' & = & \frac 1a h^{-1}k , \nonumber \\
  \gamma \longrightarrow \gamma' & = & \gamma + 
   X_q (x_0 , {\mathbf x} )\frac 1a h^{-1}k .
\label{coadaffaction}
\eea
The invariant measure on $T^*{\mathcal C}_+^{\uparrow}$ 
under this action is 
$$ 
   d\Omega_+^{\uparrow} (k , \gamma ) = 
\frac {dk_0\;d{\mathbf k}\;d\gamma_1\;d\gamma_2}{k_0^2 
   -{\mathbf k}^2}, $$
and thus the densities $\sigma_\lambda$ and $m$ appearing in 
the Wigner function (see 
(\ref{measdisintegr}), (\ref{eq:liealgmeas}) and 
(\ref{eq:explWigfcn}) become in this case, 
$$
  \sigma_+^\uparrow (k, \gamma ) = k_0^2 - {\mathbf k}^2 , 
\qquad m(\lambda , \theta ) = 
   \frac {2(\cosh\lambda - \cosh\theta )}{e^\lambda (\lambda^2
  - \theta^2 )}. $$
The final expression for the Wigner function is obtained after a 
routine computation, starting with (\ref{eq:explWigfcn}) and 
using expressions such as (\ref{matrixfcns})-(\ref{matrixfcns3}) 
in the appendix. We get, 
\bea
    W(\psi ,\;\phi\;\vert\;k , \gamma)   
    & = & \frac 1{4\pi} \int_{-\infty}^{\infty}d\lambda
                    \int_{-\infty}^{\infty}d\theta\;
           e^{-i(\gamma_1 \lambda + \gamma_2\theta )}\; 
  \overline{\psi\Bigg( 
   \frac { \sigma_3\; e^{X_q (\lambda , \theta )/2}}
  {\sinch(X_q (\lambda , \theta )/2)}k\Biggr)} \nonumber \\
       & \times & \frac {(k_0^2 -{\mathbf k}^2 )(\lambda^{2} 
     - \theta^{2})}
      {\cosh\lambda -\cosh\theta}\; \phi
     \Biggl(\frac { \sigma_3\; e^{-X_q (\lambda , \theta )/2}}
           {{\sinch}(X_q (\lambda , \theta )/2)}k \Biggr), 
\label{affWigfcnfinal}
\eea
(where we have written $\frac 1{\sinch A}$ for $[\sinch A]^{-1}$).
The above expression should be compared with (\ref{poinWigfcn}) 
and (\ref{poinWigfcn2}). Also, by virtue 
of Lemma \ref{lem:sinchinv} and (\ref{F-exp}) in the Appendix, if 
$k \in {\mathcal C}^\uparrow_+$  then so also are the arguments of 
the functions $\psi$ and $\phi$ in the above expression for the 
Wigner function. Thus we have the important result: 

\begin{thm}\label{thm:Wigfcnsupp}
  The Wigner function $W(\psi ,\;\phi\;\vert\;k , \gamma)$ in 
(\ref{affWigfcnfinal}) has support inside the coadjoint orbit,
$T^*{\mathcal C}^\uparrow_+$, associated to the representation 
$U^\uparrow_+$. 
\end{thm}

\subsection{The Weyl-Heisenberg group $\gwh$}
The Wigner function arising from the Weyl-Heisenberg group is 
the original phase space distribution introduced by Wigner 
\cite{Wig} in 1932. Although this function is well known, to 
our knowledge, it has not been obtained by the methods introduced
in this paper, linking it to the Plancherel transform.
We, therefore, give a somewhat detailed derivation
of it. Since this function was the original motivation for 
developing our 
general analysis, it is also worthwhile to put it in this 
context. (A somewhat different derivation, based on the theory 
of square integrability of  group representations,
modulo subgroups,  was given in \cite{AlAtChWo}).

   The group $\gwh$ consists of $4\times 4$ matrices 
$$ 
  g(\theta , \xi , \eta ) = \pmatrix{1 & \frac 12 \bzeta^T \omega 
          & \theta \cr {\mathbf 0} & {\mathbb I}_2 & \bzeta \cr
          0 & {\mathbf 0}^T & 1}, \qquad    \omega = \pmatrix{0 & 1\cr
          -1 & 0},   $$
with
$$  \theta \in \RR ,
          \quad \bzeta = \pmatrix{\eta \cr \xi} \in \RR^2 , 
          \quad {\mathbf 0} = \pmatrix{0 \cr 0}. $$
This group is unimodular and nilpotent. The Lie 
algebra $\bfrakg_{\rm WH}$
is generated by the three elements,
$$
  X^0 = \pmatrix{{\mathbf 0}^T & 1 \cr {\mathbb O} & 
   {\mathbf 0}}, \quad X^1 = \pmatrix{\frac 12 e_3^T &
   0 \cr {\mathbb O} & e_1}, \quad X^2  = 
   \pmatrix{-\frac 12 e_2^T & 0 \cr {\mathbb O} & e_2}, \quad
  \mathbf 0 = \pmatrix{0 \cr 0 \cr 0}, $$
$\mathbb O$ being the $3\times 3$ zero matrix and $e_1 , e_2 , 
e_3$ the canonical basis vectors in $\RR^3$. They satisfy the 
commutation relations,
$$ [X^0 , X^1 ] = [X^0 , X^2 ] = 0, \quad [X^1 , X^2 ] = X^0 . $$
Writing a 
general element of $\bfrakg_{\rm WH}$ as $Y = -\theta X^0 - 
\eta X^1 + \xi X^2$, and noting that $(Y)^2$ is the null 
matrix, we see that 
$$ 
   e^Y = g(-\theta , \xi , - \eta ) = {\mathbb I}_4 + Y . $$
Thus, the group and the Lie algebra have essentially the 
same parametrization; the Haar measure of $\gwh$ is 
$d\theta\;d\bzeta = d\theta\;d\xi \; d\eta$ and the density 
$m(X) = 1$ (see (\ref{eq:liealgmeas})), almost
everywhere. 

  Let $\{X_0^* , X_1^* , X_2^*\}$ be the dual basis 
in $\bfrakg^*_{\rm WH}$ and denote a general element in it by 
$X^* = \gamma^0 X_0^* + \gamma^1 X_1^* + \gamma^2 X_2^*$. The 
computation of the coadjoint action of $g(\theta , \xi , \eta )
\in \gwh$ on $\bfrakg^*_{\rm WH}$ 
is now a routine matter. The coordinates $\gamma^i , \;\; 
i= 1,2, 3$, transform under this action in the manner
\bea 
  \gamma^0 \longrightarrow \gamma^{0}\;' &= &\gamma^0 ,\nonumber\\
  \gamma^1 \longrightarrow \gamma^{1}\;' &= &
               \gamma_1 - \xi\gamma^0 , \nonumber \\
  \gamma^2 \longrightarrow \gamma^{2}\;' &= &\gamma_2 - \eta\gamma^0 .
\eea
The (physically) non-trivial coadjoint orbits are the planes,
$$ 
  {\mathcal O}_{\gamma_0} =  \left\{\pmatrix{\gamma_0 \cr \bgam}\;\Big|\;
         \bgam = \pmatrix{\gamma_1 \cr \gamma_2} \in \RR^2 , 
\;\; \gamma_0 \neq 0 \right\}, $$
which carry the (coad-)invariant measures 
$d\Omega_{\gamma_0} (\bgam ) = d\bgam = d\gamma_1\;d\gamma_2$, 
and comparing with (\ref{measdisintegr}) we get, 
$d\kappa (\gamma_0 ) = d\gamma_0$ and $\sigma_{\gamma_0} 
(\bgam )= 1$, 
for all $\gamma_0 \neq 0$ and almost all $\bgam = \RR^2$. 

     The (physically) non-trivial unitary irreducible representations of $\gwh$
are in one-to-one correspondence with the orbits 
${\mathcal O}_{\gamma_0}$ and thus the unitary dual 
$\widehat{\mathcal G}_{\rm WH}$ is identifiable with 
$\RR\backslash \{0\}$. We choose 
the realization,  $U_\lambda$, of the  UIR (corresponding to 
$\lambda \in \widehat{\mathcal G}_{\rm WH}$), 
which is carried by the Hilbert space 
$\h_{\lambda} = {\rm L}^2 (\RR, dx )$ and acts in the manner, 
$$ 
  (U_\lambda (\theta , \xi , \eta )\phi_\lambda )(x) = 
      e^{i\lambda\theta}\;e^{i\lambda\eta (x - \frac {\xi}2)}\;
     \phi_\lambda (x - \xi ). $$
Since $\gwh$ is unimodular, the Duflo-Moore operator $C_\lambda$ 
is a multiple of the identity, which we denote by $N_\lambda$ 
($ > 0$). Writing the Plancherel measure as $d\nu_{\rm GW} (\lambda )
= \rho (\lambda )\; d\lambda$, where $\rho$ is some density function,
we may compute both $N_\lambda$ and $\rho$ by noting that 
the orthogonality condition 
(\ref{Pl-orthog}) leads in this case to the explicit relation, 
\begin{eqnarray*}
\lefteqn{\hspace{-20pt}\frac 1{N_\lambda^2}
 \int_{\RR^3}\Bigg[\int_{\RR\backslash\{0\}} \overline{\langle 
 \psi_\lambda \vert U_{\lambda}(\theta ,\xi ,\eta)\phi_\lambda
  \rangle}\;\rho (\lambda )\;d\lambda\;\int_{\RR\backslash\{0\}} 
 \langle \psi_{\lambda'}\vert U_{\lambda'}(\theta , \xi , \eta)
 \phi_{\lambda'}\rangle}\nonumber\\
  & & \times \;\rho(\lambda')\;d\lambda'
  \Bigg]\;d\theta\; d\xi\; d\eta  
  = (2\pi)^2\int_{\RR\backslash\{0\}}\Vert\phi_\lambda\Vert^2\;
    \Vert\psi_\lambda\Vert^2\; \left[\frac {(\rho (\lambda ))^2}
      \lambda \right]\; d\lambda ,
\end{eqnarray*}
for all $\phi_\lambda , \psi_\lambda \in \h_\lambda$. We 
easily obtain, $N_\lambda = \frac 1{2\pi}$
and $\rho (\lambda ) = |\lambda|$, almost everywhere, so 
that the Plancherel measure of $\gwh$ is $|\lambda|\; d\lambda$. 

   The Wigner function is now obtained after a routine 
computation, using (\ref{eq:explWigfcn}):
\be
  W(A \;\vert\; \gamma_1 , \gamma_2 ;\; \lambda ) = 
\frac 1{(2\pi )^{\frac 12}}\int_{\RR} e^{ix\gamma_2}\;
  \overline{\psi_\lambda \left(-\frac {\gamma_1}\lambda - \frac x2 
 \right)}\;
  \phi_\lambda \left(-\frac {\gamma_1}\lambda + \frac x2 
 \right)\; dx , 
\label{standWigfcn1}
\ee
for $A \in {\mathcal B}^\oplus_2$ such that $A = 
\{\vert\phi_\lambda\rangle\langle \psi_\lambda\vert \in 
{\mathcal B}_2 (\h_\lambda )\}_{\lambda \in 
\RR\backslash\{0\}}$ and $\gamma_1 , \gamma_2 \in
{\mathcal O}_\lambda$. Note that again, in this case, the 
support of the Wigner function is concentrated on the orbit 
${\mathcal O}_\lambda$ which corresponds to the UIR $U_\lambda$.

     The above formula for the Wigner function is particularly 
interesting, since for fixed $\lambda$, 
$$ \int_{\RR^2} \vert W(A\;\vert\;\gamma_1 , \gamma_2 ;\; \lambda )
   \vert^2\; d\gamma_1\; d\gamma_2 = \vert\lambda\vert\;
   \Vert\phi_\lambda\Vert^2\;\Vert\psi_\lambda\Vert^2 . $$
This means that, for fixed $\lambda$, the expression 
(\ref{standWigfcn1}) can be used to define a function $W(A_\lambda\;
\vert\; \gamma_1 , \gamma_2 ;\; \lambda )$ on 
${\mathcal O}_\lambda \simeq \RR^2$ for any $A_\lambda \in {\mathcal B}_2
(\h_\lambda )$, such that the map 
$$ A_\lambda \longmapsto \frac 1{\vert\lambda\vert^{\frac 12}}
     W(A_\lambda \;\vert\; \cdot \; ;\; \lambda ) \in 
    {\rm L}^2 ({\mathcal O}_\lambda , d\bgam ) , $$
is an isometry. Indeed, writing 
\bea
   W_{\rm QM}(\psi_\lambda , \phi_\lambda \;\vert\; q, p ;\; \hslash )
    & = & \frac 1{(2\pi )^{\frac 12} \hslash}\;
   W\left(A_\lambda \;\Big|\; 
     -\frac q\hslash , -\frac p\hslash ;\; \frac 1\hslash 
  \right)\nonumber\\
    & = &
   \frac 1{2\pi\hslash}\int_{\RR}e^{-i\frac {xp}\hslash}
   \overline{\psi_\lambda \left(q - \frac x2 \right)}\;
     \phi\left(q + \frac x2 \right)\;dx ,
\label{standWigfcn2}
\eea
we recover the well known function originally introduced by 
Wigner \cite{Wig}. Thus, effectively, in this case Wigner 
functions can be defined for each UIR, $U_\lambda$, that appears 
in the direct integral decomposition of the regular representation 
and the support of the Wigner function is concentrated on the 
corresponding coadjoint orbit. 

\section{Conclusion}

The procedure outlined and illustrated in this paper is general 
enough to cover most groups of practical importance, for 
constructing wavelet transforms and Wigner functions. There are 
still, however, group representations which are used in 
practical applications, but which are not 
amenable to the present technique. Representations which are not
in the support of the Plancherel measure fall into this category. 
For example, in the case of the two Poincar\'e groups discussed 
here, the representations corresponding to the boundaries of 
the cones (the ``mass zero representations'') fall outside of our 
scheme. One interesting direction for further research concerns 
the use of adapted Fourier transforms for the construction of
decomposable Wigner maps. Also, the precise relationship between
Wigner maps and deformation quantization (which is where the
adapted Fourier transforms originate) should be worked out
explicitly.

\section*{Acknowledgements}

We would like to thank M. Cahen and S. Gutt for stimulating
discussions and pointing out various references to us.
The authors would like to acknowledge 
financial support from the Natural Sciences and Engineering 
Research Council (NSERC), Canada and the Fonds pour 
la Formation de Chercheurs et l'Aide \`a la Recherche 
(FCAR), Qu\'ebec. HF would like to thank the Department
of Mathematics and Statistics of Concordia University,
Montr\'eal, for their hospitality. STA would also like to 
thank G. Schlichting for hospitality at the Zentrum Mathematik der 
Technischen Universit\"at M\"unchen, where part of this work was 
completed.

\section*{Appendix}
  We collect in this appendix a few formulae and 
results for the 
for the various special matrix functions which appear in 
this paper. We begin by defining three real 
valued functions on 
$\RR$:
\bea 
  \sinch x & := & 1 + \frac{x^2}{3!} + \frac{x^4}{5!} 
      + \frac{x^6}{7!} + \ldots \nonumber\\
           & = & \frac{\sinh x}x , 
  \quad \mbox{\rm if $x \neq 0$}, \label{sinch}\\
  \cosinch x & := & \frac{x}{2!} + \frac{x^3}{4!} 
      + \frac{x^5}{6!} + \frac{x^7}{8!} + \ldots  \nonumber\\
            & = & \frac{\cosh x - 1}x, \quad 
           \mbox{\rm if $x \neq 0$}, \label{cosinch}\\
  F(x) & := & \sinch x + \cosinch x  = 
           e^{\frac x2}\;\sinch (\frac x2 ) 
        = 1 + \frac{x}{2!} + \frac{x^2}{3!} + 
              \frac{x^3}{4!} + \ldots \nonumber\\
        & = & \frac {e^x - 1}x , \quad 
                     \mbox{\rm if $x \neq 0$} \label{F-fcn}.
\eea
Note that $\sinch x $ is an even function of $x$, while 
$\cosinch x$ is an 
odd function. The inverse of $F(x)$ has an interesting 
series expansion:
\bea F(x )^{-1} & = & e^{-x}F(-x)^{-1}
 = \frac {e^{-\frac x2}}{\sinch (\frac x2 )} 
     =  1 - \frac {x}{2} + \sum_{k\geq 1} 
        (-1)^{k-1} \;\frac {B_k x^{2k}}{(2k)!} \nonumber \\
    & = &  \frac x{e^x - 1}, \quad \mbox{\rm if $x \neq 0$} 
\label{F-invfcn}
\eea
where the $B_k$ are the Bernoulli numbers, $B_1 = \frac 16 , \;
B_2 = \frac 1{30} , \; B_3 = \frac 1{42}, \; B_4 = \frac 1{30}$,
etc., and generally, 
$$ B_k = \frac {(2k)!}{\pi^{2k}\; 2^{2k-1}}\sum_{n=1}^{\infty}
     \frac 1{n^{2k}} \;.$$ 
     
   If $A$ is an $n\times n$ matrix, then using the 
series expansions, we can define 
the matrix versions $\sinch A ,\; \cosinch A , \; F(A)$ 
of these functions. In addition, 
if $\mbox{\rm det}\; A  \neq 0$, then we can also write, 
$\sinch A = A^{-1}\sinh A, \;\;
\cosinch A = A^{-1}(\cosh A - 1)$, etc. In particular, 
for the matrix
\be
   X_q (\lambda , \theta ) = \lambda {\mathbb I}_2 +
  \theta \sigma_1 , \quad \lambda , \theta \in \RR , 
     \quad  {\mathbb I}_2 = \pmatrix{1 & 0 \cr 0 & 1}, 
\quad \sigma_1 = \pmatrix{0 & 1 \cr 1 & 0}, 
\label{Liematrix}
\ee
we easily compute, 
\bea
  e^{X_q (\lambda , \theta )} & = & 
        e^\lambda \cosh\theta\; {\mathbb I}_2 + 
             e^\lambda \sinh\theta\; \sigma_1 , 
     \qquad \mbox{\rm det}\;[e^{X_q (\lambda , \theta )}]
           = e^{2\lambda} , \nonumber \\
F(X_q (\lambda , \theta ))& = & e^{\frac {X_q (\lambda , \theta )}2}
    \;\sinch \left(\frac {X_q (\lambda , \theta )}2 \right), 
   \label{F-exp} \\ 
    \mbox{\rm det}\;[F(X_q (\lambda , \theta ))] & = & 
   \frac {2e^\lambda (\cosh\lambda - \sinh\theta )}
       {\lambda^2 - \theta^2 } , \quad 
   \mbox{\rm det}\;[F(X_q (0 , \theta ))] 
    = \sinch^2 (\frac \theta 2 ), \nonumber \\
  \sinch X_q (0, \theta ) & = &\sinch\theta \;{\mathbb I}_2, \qquad 
    \cosinch X_q (0, \theta )= \cosinch\theta\;\sigma_1 , \nonumber\\
  F(X_q (0, \theta )) & = & \sinch\theta \;{\mathbb I}_2 + 
             \cosinch\theta\; \sigma_1 = 
      \sinch (\frac \theta 2 )\; e^{\frac {X_q (0, \theta )}2}, 
\label{matrixfcns}
\eea
To these we 
add the useful relationships, 
\bea 
  \sigma_3 \; e^{X_q (\lambda , \theta )}\;\sigma_3 & 
   = & e^{X_q (\lambda , -\theta )}, \qquad 
\sigma_3 \; F(X_q (\lambda , \theta ))\; \sigma_3 
  = F(X_q (\lambda , -\theta )), \nonumber \\
\sigma_1 \; e^{X_q (\lambda , \theta )}\;\sigma_1 
   & = & e^{X_q (\lambda , \theta )},
 \qquad \sigma_3 
   =  \pmatrix{1 & 0 \cr 0 & -1}, \quad \sigma_1^2 
      = \sigma_3^2 = {\mathbb I}_2 .
\label{matrixfcns2}
\eea

   Note that the matrices $X_q (\lambda , \theta )$ span the 
Lie algebra of the $SO_{\rm Aff}(1,1)$, which is 
the group $SO_0 (1,1)$ together with dilations. Let 
us denote this Lie algebra by 
${\mathfrak s}{\mathfrak o}_{\rm Aff}(1,1)$ and note that  
the group $SO_{\rm Aff}(1,1)$
itself consists of all invertible elements of 
this algebra. The set ${\mathfrak s}{\mathfrak o}_{\rm Aff}(1,1)$
is closed under ordinary matrix multiplication and under this 
multiplication it is a commutative set. Furthermore, the  matrices 
$F(X_q (\lambda , \theta ))$ are elements of 
${\mathfrak s}{\mathfrak o}_{\rm Aff}(1,1)$, for all 
$\lambda , \theta \in \RR$.  For any two vectors, $k = 
\pmatrix{k_0 \cr {\mathbf k}}, \;\; u = 
\pmatrix{u_0 \cr {\mathbf u}}\in \RR^2$, and any $2\times 2$
matrix $A$, 
\bea
  X_q (u_0 , {\mathbf u} )k & = & X_q (k_0 , \mathbf k )u , 
\nonumber \\ 
    \langle A k \; , \sigma_3 A u \rangle  & = &
   \mbox{\rm det}\;A\;\langle k \; , u \rangle = 
\mbox{\rm det}\;A\;(k_0 u_0 - {\mathbf k}{\mathbf u} ).
\label{matrixfcns3}
\eea

  Diagonalizing the matrices $X_q (\lambda , \theta )$, 
 $$ 
    VX_q (\lambda , \theta )V^T = \pmatrix{\lambda + \theta & 0 \cr 
                                  0 & \lambda - \theta}, \qquad 
          V = \frac 1{\sqrt{2}}\pmatrix{1 & 1 \cr -1 & 1}, $$
we get, 
\bea
V F(X_q (\lambda , \theta )) V^T & = &\pmatrix{F(\lambda + \theta )& 0 \cr 
                                  0 & F(\lambda - \theta )}, \nonumber\\ 
        \mbox{\rm det}\;[F(X_q (\lambda , \theta ))] & = &
           F(\lambda + \theta )\;F(\lambda - \theta ) > 0 .  
\label{diagmatrixeqn}
\eea
Writing 
 $$ \zeta = \pmatrix{\zeta_1 \cr \zeta_2} = V\pmatrix{k_0 \cr \mathbf k } = 
Vk,$$
we see that the condition that $k \in {\mathcal C}^\uparrow_+$  (i.e., 
$k_0^2 > {\mathbf k}^2 , \;\; k_0 > 0$), is equivalent to 
having $\zeta_1 , 
\zeta_2 > 0$. Thus we have the result,

\begin{lemma}\label{lem:sinchinv}
If $k \in {\mathcal C}^\uparrow_+$, then $F(X_q (\lambda , \theta ))k \in 
{\mathcal C}^\uparrow_+$, for all $\lambda , \theta \in \RR$.
\end{lemma}
\begin{prf} 
By (\ref{diagmatrixeqn}), 
$$  VF(X_q (\lambda , \theta ))V^T \zeta = 
   \pmatrix{F(\lambda + \theta )\zeta_1 
       \cr F(\lambda - \theta )\zeta_2 }, $$
and since both  $F(\lambda + \theta ), 
F(\lambda - \theta ) > 0$, the condition $\zeta_1 , 
\zeta_2 > 0$ is preserved under the 
action of $VF(X_q (\lambda , \theta ))V^T $. Hence the condition
$k \in {\mathcal C}^\uparrow_+$ is preserved under the action of 
$F(X_q (\lambda , \theta ))$. 
\end{prf}


\end{document}